\documentclass[prd,epsf,aps,twocolumn]{revtex4}
\begin{document}
\title{Gravitational Field of Massive Point Particle in General Relativity}
\author{P.~P.~Fiziev\footnote{ E-mail:\,\, fiziev@phys.uni-sofia.bg} }
\affiliation{Department of Theoretical Physics, Faculty of
Physics, Sofia University, 5 James Bourchier Boulevard,
Sofia~1164, Bulgaria.\\and\\ The Abdus Salam International  Centre
for Theoretical Physics, Strada Costiera 11, 34014 Trieste, Italy.
}
\begin{abstract}
Utilizing various gauges of the radial coordinate we give a
description of static spherically symmetric space-times with
point singularity at the center and vacuum outside the
singularity. We show that in general relativity (GR) there exist
a two-parameters family of such solutions to the Einstein equations 
which are physically distinguishable but only some of them describe 
the gravitational field of a single massive point particle with
nonzero bare mass $M_0$. In particular, we show that the widespread
Hilbert's form of Schwarzschild solution, which depends only on the 
Keplerian mass $M<M_0$, does not solve the Einstein equations with a massive 
point particle's stress-energy tensor as a source. 
Novel normal coordinates for the field and a new physical
class of gauges are proposed, in this way achieving a correct
description of a point mass source in GR. We also introduce a
gravitational mass defect of a point particle and determine the
dependence of the solutions on this mass defect. 
The result can be described as a change of the Newton potential
$\varphi_{\!{}_N}=-G_{\!{}_N}M/r$ to a modified one: 
$\varphi_{\!{}_G}=-G_{\!{}_N}M/
\left(r+G_{\!{}_N} M/c^2\ln{{M_0}\over M}\right)$ 
and a corresponding modification of the four-interval.  
In addition we
give invariant characteristics of the physically and geometrically
different classes of spherically symmetric static space-times
created by one point mass. These space-times are analytic
manifolds with a definite singularity at the place of the matter
particle.

\noindent{PACS number(s): 04.20.Cv, 04.20.Jb, 04.20.Dw}
\end{abstract}
%
\sloppy
\newcommand{\lfrac}[2]{{#1}/{#2}}
\newcommand{\sfrac}[2]{{\small \hbox{${\frac {#1} {#2}}$}}}
\newcommand{\ben}{\begin{eqnarray}}
\newcommand{\een}{\end{eqnarray}}
\newcommand{\la}{\label}
\maketitle
%
\section{Introduction}

The Einstein's equations:
\ben G^\mu_\nu = \kappa
T^\mu_\nu \la{Einst}\een determine the solution of a given
physical problem up to four arbitrary functions, i.e., up to a
choice of coordinates. This reflects the well known fact that GR
is a gauge theory.

According to the standard textbooks
\cite{books} the fixing of the gauge in GR in a {\em holonomic}
frame is represented by a proper choice of the quantities
\ben\bar\Gamma_\mu\!=\!-{{1}\over{\sqrt{|g|}}}g_{\mu\nu}
\partial_\lambda\left(\sqrt{|g|}g^{\lambda\nu}\right),\la{Gammas}\een which 
emerge
when one expresses the 4D d'Alembert operator in the form
$g^{\mu\nu}\nabla_\mu\nabla_\nu=g^{\mu\nu}
\left(\partial_\mu\partial_\nu-\bar\Gamma_\mu\partial_\nu\right)$.

We shall call the change of the gauge fixing expressions (\ref{Gammas}),
{\em without} any preliminary conditions on the analytical behavior
of the used functions, a gauge transformations in a {\em broad} sense.
This way we essentially expand the class of
admissible gauge transformations in GR. Of course this will alter
some of the well known {\em mathematical} results in the commonly
used mathematical scheme of GR.
We think that a careful analysis of such wider framework for the gauge
transformations in GR can help us clarify some long standing {\em physical}
problems. After all, in the physical applications
the mathematical constructions are to reflect in an
adequate way the properties of the real physical objects.

It is well known that in the gauge theories we may have different
solutions with the same symmetry in the base space. Such solutions
belong to different gauge sectors, owning different geometrical,
topological and physical properties. For a given solution there
exists a class of {\em regular} gauges, which alter the form of the
solution without changing these essential properties. In contrast,
by performing a {\em singular} gauge transformation one can change
both the geometrical, topological and physical properties of the
solution, making a transition to another gauge sector.

In geometrical sense the different solutions in GR
define different 4D pseudo-Riemannian space-time manifolds
${\cal M}^{(1,3)}\{g_{\mu\nu}(x)\}$.
A subtle point in GR formalism is that transitions from a given physical
solution to an essentially different one, can be represented as a "change of
coordinates". This is possible because from gauge theory point of
view the choice of space-time coordinates in GR means simultaneously two
different things: choosing a gauge sector in which the solution
lives and at the same time fixing the (regular) gauge in this sector.
While the latter solves an inessential local gauge fixing problem,
the first fixes the essential global properties of the solution.

One can understand better this peculiarity of GR in the framework
of its modern differential geometrical description as a gauge
theory, using the principal frame bundle $F\left({\cal
M}^{(1,3)}\{g_{\mu\nu}(x)\}\right)$. (See, for example, the second
edition of the monograph  \cite{Exact}, \cite{GS}, and the
references therein.) The change of coordinates on the base
space ${\cal M}^{(1,3)}\{g_{\mu\nu}(x)\}$ induces {\em
automatically} a nontrivial change of the frames on the fibre of
frames. Singular coordinate transformations may produce a
change of the gauge sector of the solution, because they may
change the topology of the frame bundle, adding new singular
points and(or) singular sub-manifolds, or removing some of the
existing ones. For example, such transformations can alter the
fundamental group of the base space ${\cal
M}^{(1,3)}\{g_{\mu\nu}(x)\}$, the holonomy group of the affine
connection on
$F\left({\cal M}^{(1,3)}\{g_{\mu\nu}(x)\}\right)$, etc.
Thus one sees that coordinate changes
in broad sense are more than a pure alteration of the
labels of space-time points.

If one works in the framework of the theory of {\em smooth real}
manifolds, ignoring the analytical properties of the solutions
in the complex domain, one is generally allowed to change the gauge
without an alteration of the physical problem in the {\em real}
domain, i.e. without change of the boundary conditions,
as well as without introduction of new singular points,
or change of the character of the existing ones.
Such special type of {\em regular}
gauge transformations in GR describe the diffeomorphisms of
the {\em real} manifold ${\cal M}^{(1,3)}\{g_{\mu\nu}(x)\}$.
This manifold is already fixed by the initial choice of the gauge.
Hence, the {\em real} manifold ${\cal M}^{(1,3)}\{g_{\mu\nu}(x)\}$
is actually described by a class of equivalent gauges, which correspond
to all diffeomorphisms of this manifold and are related with regular gauge
transformations.

The transitions between some specific real manifolds ${\cal
M}^{(3)}\{-g_{mn}({\bf r })\}$, which are not diffeomorphic, can be
produced by the use of proper {\em singular} gauge
transformations. If we start from an everywhere smooth manifold,
after a singular transformation we will have a new manifold with
some singularities, which are to describe real physical
phenomena in the problem at hand. The inclusion of such singular
transformations in our consideration yields the necessity to talk
about gauge transformations {\em in a broad sense}. They are
excluded from present-day standard considerations by the commonly
used assumption, that in GR one has to allow {\em only}
diffeomorphic mappings.

Similar singular transformations are well known in gauge theories
of other fundamental physical interactions: electromagnetic
interactions, electroweak interactions, chromodynamics.
For example, in the gauge theories singular gauge
transformations describe transitions between solutions in
topologically different gauge sectors.
Singular gauge transformations are used in the theory of
Dirac monopole, vortex solutions, t'Hooft-Poliakov monopoles,
Yang-Mills instantons, etc.
See, for example, \cite{Rubakov, GS} and the references therein.

The simplest example is the singular gauge transformation of the
3D vector potential: ${\bf A}({\bf r}) \to {\bf A}({\bf r})+\nabla
\varphi({\bf r})$ in electrodynamics, defined in Cartesian
coordinates $\{x,y,z\}$ by the singular gauge function
$\varphi=\alpha \arctan(y/x)$ ($\alpha=const$). Suppose that
before the transformation we have had a 3D space ${\cal
M}^{(3)}\{-g_{mn}({\bf r})=\delta_{mn}\}={\cal
R}^{(3)}\{\delta_{mn}\}$. Then this singular gauge transformation
removes the whole axes $OZ$ out of the Euclidean 3D space ${\cal
R}^{(3)}\{\delta_{\mu\nu}\}$, changing the topology of this part of
the base space. As a result the quantity $\oint_Cd{\bf r} {\bf
A}({\bf r})$, which is gauge invariant under regular gauge
transformations, now changes its value to $\oint_Cd{\bf r} {\bf
A}({\bf r})+2\pi N \alpha$, where $N$ is the winding number of the
cycle $C$ around the axes $OZ$. Under such singular gauge
transformation the solution of some initial physical problem will
be transformed onto a solution of a completely different problem.

At present the role of singular gauge transformations in the above
physical theories is well understood.

In contrast, we still don't have systematic study of the classes
of physically, geometrically and topologically different solutions
in GR, created by singular gauge transformations, even in the simple
case of static spherically symmetric space-times with only one point
singularity, although the first solution of this type was discovered
first by Schwarzschild nearly 90 years ago \cite{Schwarzschild}.
Moreover, in GR at present there is no clear understanding both of the
above gauge problem and of its physical significance.

Here we present some initial steps toward the clarification of the role
of different GR gauges in broad sense for spherically symmetric
static space-times with point singularity at the center of symmetry
and vacuum outside this singularity.

\section{Static Spherically Symmetric Space-Times with Point Source of 
Gravity}

The static point particle with bare mechanical rest mass $M_0$
can be treated as a 3D entity. Its proper frame of reference is
most suitable for description of the {\em static} space-time with this
single particle in it. We prefer to present the problem of a point source
of gravity in GR as an 1D mathematical problem, considering the
dependence of the corresponding functions on the only essential
variable -- the radial variable $r$. This can be achieved in
the following way.

The spherical symmetry of the 3D space reflects adequately
the point character of the source of gravity.
A real spherically symmetric 3D Riemannian space
${\cal M}^{(3)}\{-g_{mn}({\bf r })\} \subset {\cal 
M}^{(1,3)}\{g_{\mu\nu}(x)\}$
can be described using standard spherical coordinates
$r\in[0,\infty),\,\theta\in[0,\pi),\phi\in[0,2\pi)$. Then $${\bf
r}=\{x^1,x^2,x^3\}=\{r\sin\theta\cos\phi,r\sin\theta\sin\phi,r\cos\theta\}.$$

The physical and geometrical meaning of the radial coordinate $r$
is not defined by the spherical symmetry of the problem and is
unknown \textit{a priori} \cite{Eddington}. The only clear thing
is that its value $r=0$ corresponds to the center of the symmetry,
where one must place the physical source of the gravitational
field. In the present article we assume that there do not exist
other sources of gravitational field outside the center of
symmetry.

There exists unambiguous choice of a global time $t$ due to the
requirement to use a static metric. In proper units it yields the
familiar form of the space-time interval \cite{ComJann}:
\ben
ds^2\!=\!g_{tt}(r)\,dt^2\!+\!g_{rr}(r)\,dr^2\!-\!\rho(r)^2(d\theta^2\!+
\!\sin^2\theta\,d\phi^2) \la{ds0}\een
with unknown functions $g_{tt}(r)>0,\,g_{rr}(r)<0,\,\rho(r)$.

In contrast to the variable $r$, the quantity $\rho$ has a clear
geometrical and physical meaning: It is well known that $\rho$
defines the area $A_\rho=4\pi\rho^2$ of a centered at $r=0$ sphere
with "a radius" $\rho$ and the length of a big circle on it
$l_\rho=2\pi\rho$. One can refer to this quantity as an {\em ``area 
radius''},
or as an optical
{\em "luminosity distance``}, because the luminosity of
distant physical objects is reciprocal to $A_\rho$.

One has to stress that in the proper frame of reference of the point 
particle  this quantity does not measure the {\em real} geometrical 
distances in the corresponding curved space-time. In contrast, if 
$\rho_{fixed}$ is some arbitrary fixed value of the luminosity distance,
the expression 
$\left(\rho_2-\rho_1\right)/\left( 1-2M/\rho_{fixed}\right)^{1/2}$
measures the 3D geometrical distance between the geometrical points 
$2$ and $1$ on a radial geodesic line in the frame of free falling 
clocks \cite{GH}. 
Nevertheless, even in this frame the absolute value of the variable
$\rho$ remains not fixed by the 3D distance measurements.

In the static spherically symmetric case
the choice of spherical coordinates and static metric
dictates the form of three of the gauge fixing
coefficients (\ref{Gammas}):\,
$\bar\Gamma_t\!=\!0,\,\,\bar\Gamma_\theta\!=
-\!\cot\theta,\,\,\bar\Gamma_\phi\!=\!0$,
but the form of the quantity
$\bar\Gamma_r\!=\left(\!
\ln\left({\sqrt{-g_{rr}}\over{\sqrt{g_{tt}}\,\rho^2}}\right)\!\right)^\prime$,
and, equivalently, the function $\rho(r)$
are still not fixed. Here and further on, the prime denotes
differentiation with respect to the variable $r$. We refer to the
freedom of choice of the function $\rho(r)$ as "{\em a rho-gauge
freedom}", and to the choice of the $\rho(r)$ function as
"{\em a rho-gauge fixing}" in a {\em broad} sense.

In the present article we will not use more general gauge
transformations in a broad sense, than the rho-gauge ones.
In our 1D approach to the problem at hand all possible
mathematical complications, due to the use of such wide class
of transformations, can be easily controlled.

The overall action for the aggregate of a point particle and its
gravitational field is ${\cal A}_{tot}\!=\!{\cal A}_{GR}+{\cal
A}_{M_{0}}$. Neglecting the surface terms one can represent the
Hilbert-Einstein action ${\cal A}_{GR}\!=\!-{1\over{16\pi
G_n}}\int\!d^4x\sqrt{|g|}R$ and the mechanical action ${\cal
A}_{M_{0}}\!=\!-M_0\int\!ds$ of the point source with a bare mass
$M_0$ as integrals with respect to the time $t$ and the radial
variable $r$ of the following Lagrangian densities: \ben {\cal
L}_{GR}={\frac {1} {2 G_N}}\!\left({
{2\rho\rho^\prime\left(\sqrt{g_{tt}}\right)^\prime\!+\!
\left(\rho^\prime\right)^2\sqrt{g_{tt}} } \over {\sqrt{\!-g_{rr}}}
}\!+\!\sqrt{g_{tt}}\sqrt{\!-g_{rr}}\right),\nonumber\\ {\cal
L}_{M_{0}}= - M_0\sqrt{g_{tt}}\delta(r).\hskip 4.85truecm
\la{LGR}\een Here $G_N$ is the Newton gravitational constant,
$\delta(r)$ is the 1D Dirac function \cite{Gelfand}. (We are using
units $c\!=\!1$.)

As a result of the rho-gauge freedom the field variable
$\sqrt{\!-g_{rr}}$ is not a true dynamical variable but rather
plays the role of a Lagrange multiplier, which is needed in a
description of a constrained dynamics. This auxiliary variable
enters the Lagrangian ${\cal L}_{GR}$ in {\em a nonlinear} manner,
in a contrast to the case of the standard Lagrange multipliers.
The corresponding Euler-Lagrange equations read: \ben
\left({{2\rho\rho^\prime}\over{\sqrt{\!-g_{rr}}}}\right)^\prime
-{{{\rho^\prime}^2}\over{\sqrt{\!-g_{rr}}}}-\sqrt{\!-g_{rr}}+2 G_N
M_0 \delta(r)=0,
\nonumber\\
\left({{\left(\rho\sqrt{\!g_{tt}}\right)^\prime}\over{\sqrt{\!-g_{rr}}}}
\right)^\prime
-{{\rho^\prime\left(\sqrt{\!g_{tt}}\right)^\prime}\over{\sqrt{\!-g_{rr}}}}=0,
\nonumber\\
{{2\rho\rho^\prime\left(\sqrt{g_{tt}}\right)^\prime\!+\!
\left(\rho^\prime\right)^2\sqrt{g_{tt}} } \over {\sqrt{\!-g_{rr}}}
}\!-\!\sqrt{g_{tt}}\sqrt{\!-g_{rr}}\stackrel{w}{=}0 \la{EL}\een
where the symbol "\,$\stackrel{w}{=}$\," denotes a weak equality
in the sense of the theory of constrained dynamical systems.

If one ignores the point source of the gravitational field, thus
considering only the domain $r>0$ where $ \delta(r)\equiv 0$, one
obtains the standard solution of this system \cite{ComJann}: \ben
g_{tt}(r)=1-{{\rho_G}/{\rho(r)}},\,\,\, g_{tt}(r)\, g_{rr}(r)=
-\left(\rho(r)^\prime\right)^2, \la{ss}\een where $\rho_G=2G_N M$
is the Schwarzschild radius, $M$ is the gravitational
(Keplerian) mass of the source, and $\rho(r)$ is 
an arbitrary ${\cal C}^1$ function.

\section{Some Examples of Different Radial Gauges}

In the literature one can find different choices of the function
$\rho(r)$ for the problem at hand:

1. Schwarzschild gauge \cite{Schwarzschild}:
$\rho(r)\!=\!\left(r^3\!+\!\rho_G^3\right)^{1/3}$. It produces
$\bar\Gamma_r\!=\!-{2\over {r}}
\left(\rho(r^3\!-\!\rho_G^3)\!-\!\rho_G(r^3/2\!-\!\rho_G^3)\right)\!/\!
\left(\rho^3(\rho\!-\!\rho_G)\right)$.

2. Hilbert gauge \cite{Hilbert}: $\rho(r)=r$. It gives
$\bar\Gamma_r=-{2\over {r}}\left(1-{{\rho_G}\over
{2r}}\right)/\left(1-{{\rho_G}\over {r}}\right)$. This simple
choice of the function $\rho(r)$ is often related incorrectly with
the original Schwarzschild article \cite{Schwarzschild} (See
\cite{AL}).  In this case the coordinate $r$ coincides with the
luminosity distance $\rho$ and the physical domain $r\in
[0,\infty)$ contains an event horizon at $\rho_{{}_H}=\rho_G$.
This unusual circumstance forces one to develop a nontrivial
theory of black holes for Hilbert gauge -- see for example
\cite{books,FN} and the references therein.

3. Droste gauge \cite{Droste}:  The function $\rho(r)$ is given
implicitly by the relations $\rho/\rho_G=\cosh^2\psi\geq 1$ and
$r/\rho_G=\psi+\sinh\psi\cosh\psi$. The coefficient
$\bar\Gamma_r=-{2\over\rho}\left({\rho\over{\rho_G}}-{3\over 4}
{{\rho_G}\over{\rho}}\right)/ \sqrt{1-{{\rho_G}\over{\rho}}}$. For
this solution the variable $r$ has a clear geometrical meaning: it
measures the 3D-radial distance to the center of the spherical
symmetry.

4. Weyl gauge \cite{Weyl}: $\rho(r)\!=\!{{1}\over
4}\!\left(\!\sqrt{r/{\rho_G}}\!+\!\sqrt{{\rho_G}/
r}\right)^2\!\geq\!\rho_G$ and $\bar\Gamma_r\!=\!-{2\over
{r}}/\left(1\!-\!{{\rho_G^2}\over{r^2}}\right)\!=\!-{2\over
{\rho_G}}/\left({{r}\over{\rho_G}}\!-\!{{\rho_G}\over{r}}\right)$.
In this gauge the 3D-space ${\cal M}^{(3)}\{-g_{mn}({\bf r })\}$
becomes obviously conformally flat. The coordinate $r$ is the
radial variable in the corresponding Euclidean 3D space.

5. Einstein-Rosen gauge \cite{Einstein}. In the original article
\cite{Einstein} the variable $u^2=\rho-\rho_G$ has been used. To
have a proper dimension we replace it by $r=u^2\geq 0$. Hence,
$\rho(r)=r+\rho_G\geq \rho_G$ and $\bar\Gamma_r=-{2\over
{r}}{{r+\rho_G/2}\over {r+\rho_G}}$.

6. Isotropic t-r gauge, defined according to the formula
$r\!=\!\rho\!+\!\rho_G\ln\left({{\rho}\over{\rho_G}}\!-\!1\right)$,
$\rho\!\geq\!\rho_G$. Then only the combination $(dt^2\!-\!dr^2)$
appears in the 4D-interval $ds^2$ and
$\bar\Gamma_r\!=\!-{2\over\rho}\left(1\!-\!{{\rho_G}\over{\rho}}\right)$.

7. One more rho-gauge  in GR was introduced by Pugachev and Gun'ko
\cite{ComJann} and, independently, by Menzel \cite{Menzel}. In
comparison with the previous ones it is simple and more natural
with respect to the quantity $\bar\Gamma_r$. We fix it using the
condition $\bar\Gamma_r\!=\!-{2\over r}$ which is identical with
the rho-gauge fixing for spherical coordinates in a flat
space-time. Thus the curved space-time coordinates $t, r,
\theta,\phi$ are fixed in a complete {\em coherent} way with the
flat space-time spherical ones.

Then, absorbing in the coordinates' units two inessential
constants, we obtain a novel form of the 4D interval for a point
source: $$ ds^2 \!=\! e^{2\varphi_{\!{}_N}({\bf
r})}\left(\!dt^2\!-\!{{dr^2}\over{N(r)^4}}\!\right)\!-\!{{r^2}\over{N(r)^2}}
\left(d\theta^2\!+\!sin^2\theta d\phi^2\right),$$ where
$N(r)\!=\!\left(2\varphi_{\!{}_N}\right)^{-1}
\left(e^{2\varphi_{\!{}_N}}-1\right)$; 
$N(r)\!\sim\!1\!+\!{\cal O}({{\rho_G}\over r})$ --
for $r\!\to\!\infty$ and $N(r)\!\sim\!{{r}\over \rho_G}$ -- for
$r\!\to\! +0$. In this specific gauge $\rho(r)\!=\!r/N(r)$ and
$\rho(r)\!\sim\!r$ -- for $r\!\to\!\infty$, $\rho(r)\!\to\!\rho_G$
-- for $r\!\to\!+0$.

The most remarkable property of this flat space-time-coherent
gauge is the role of the {\em exact} classical Newton
gravitational potential $\varphi_{\!{}_N} ({\bf r})=-{{G_N M}\over
r}$ in the above {\em exact} GR solution. As usual, the component
$g_{tt}=e^{2\varphi_{\!{}_N} ({\bf r})}\sim
1+2\varphi_{\!{}_N}({\bf r})+{\cal O}(\varphi_{\!{}_N}({\bf r}))$
-- for $r\to \infty$. But now
$g_{tt}$ has an essentially singular point at $r=0$.

The new form of the metric is asymptotically flat for
$r\to\infty$. For $r\to+0$ it has a limit  $d\theta^2+sin^2\theta
d\phi^2$. The geometry is regular in the whole space-time ${\cal
M}^{(1,3)}(g_{\mu\nu}(x))$ because the two nonzero scalars
${1\over {48}} R_{\mu\nu\lambda\kappa}
R^{\mu\nu\lambda\kappa}\!=\!{1\over
4}\left(1\!-\!g_{00}\right)^4\!=\!{1\over
4}\left(1\!-\!e^{2\varphi_{\!{}_N}({\bf r})}\right)^4$ and
${1\over {96}} R_{\mu\nu\lambda\kappa}
R^{\lambda\kappa}{}_{\sigma\tau}
R^{\sigma\tau\mu\nu}\!=\!-\!{1\over 8}\left(1\!-\!g_{00}\right)^6
\!=\!-{1\over 8}\left(1\!-\!e^{2\varphi_{\!{}_N}({\bf
r})}\right)^6$ are finite everywhere, including the center
$r\!=\!+0$. At this center we have a zero 4D volume, because
$\sqrt{|g|}=e^{4\varphi_{\!{}_N}({\bf r})}N(r)^{-8}
r^2\sin\theta$. At the same time the coordinates
$x^m\in(-\infty,\infty), \,m=1,2,3;$ are globally defined in
${\cal M}^{(1,3)}(g_{\mu\nu}(x))$. One can check that $\Delta_g
x^m=2\left({{N}\over r}\right)^3 x^m$. Hence, $\Delta_g x^m\sim
{\cal O}\left({1\over {r^2}}\right)$ -- for $r\to \infty$ and
$\Delta_g x^m\sim 2 x^m\sim {\cal O}\left({r}\right)$ -- for $r\to
+0$, i.e., the coherent coordinates $x^m$ are asymptotically
harmonic in both limits, but not for finite values of
$r\approx \rho_G$.

At this point one has to analyze two apparent facts:

i) An event horizon $\rho_{{}_H}$ exists in the physical domain
{\em only} under Hilbert choice of the function $\rho(r)\equiv r$,
but not in the other gauges, discussed above. This demonstrates
that the existence of black holes strongly depends on this
choice of the rho-gauge in a {\em broad} sense.

ii) The choice of the function $\rho(r)$ can change drastically
the character of the singularity at the place of the point source
of the metric field in GR.

New interesting gauge conditions for the radial variable $r$ and
corresponding gauge transformations were introduced and
investigated in \cite{Bel1}. In the recent articles \cite{Bel2} it
was shown that using a nonstandard $\rho$-gauge for the
applications of GR to the stelar physics one can describe extreme
objects with {\em arbitrary large} mass, density and size. In
particular, one is able to shift the value of the
Openheimer-Volkoff limiting mass $0.7 M_\odot$ of neutron stars to
a new one: $3.8 M_\odot$. This may shed a new light on these
astrophysical problems and once more shows that the choice of
rho-gauge (i.e. the choice of the coordinate $r$) can have
real physical consequences.

Thus we see that by analogy with the classical electrodynamics and
non-abelian gauge theories of general type, in GR  we must use
indeed a more refined terminology and corresponding mathematical
constructions. We already call the choice of the function
$\rho(r)$ "a rho-gauge fixing in a broad sense". Now we see that
the different functions $\rho(r)$ may describe {\em different}
physical solutions of Einstein equations (\ref{Einst}) with the
{\em  same} spherical symmetry in presence of only one singular
point at the center of symmetry. As we have seen, the mathematical
properties of the singular point may depend on the choice of the
rho-gauge in a broad sense. Now the problem is to clarify the
physical meaning of the singular points with different
mathematical characteristics.

Some of the static spherically symmetric solutions with a
singularity at the center can be related by regular gauge
transformations. For example, choosing the Hilbert  gauge we
obtain a black hole solution of the {\em vacuum} Einstein
equations (\ref{Einst}). Then we can perform a {\em regular} gauge
transformation to harmonic coordinates, defined by the relations
$\bar\Gamma_t=0$, $\bar\Gamma_{x^1}=0$, $\bar\Gamma_{x^2}=0$,
$\bar\Gamma_{x^3}=0$, preserving the existence of the event
horizon, i.e. staying in the same gauge sector. In the next
sections we will find a geometrical criteria for answering the
question: "When the different representations of
static spherically symmetric solutions with
point singularity describe diffeomorphic space-times ?"

The definition of the regular gauge transformations allows
transformations that do not change the number and the character of
singular points of the solution in the real domain and the
behavior of the solution at the boundary, but can place these
points and the boundary at new positions.

\section{General Form of the Field Equations in Hilbert Gauge
and some Basic Properties of Their Solutions}

According to Lichnerowicz \cite{Lichnerowicz, YB, Exact}, the
physical space-times ${\cal M}^{(1,3)}\{g_{\mu\nu}(x)\}$ of
general type must be smooth manifolds at least of class $C^2$ and
the metric coefficients $g_{\mu\nu}(x)$ have to be at least of
class $C^3$, i.e. at least three times continuously differentiable
functions of the coordinates $x$. When one considers $g_{tt}(r)$,
$g_{rr}(r)$ and $\rho(r)$ not as distributions, but as usual
functions of this class, one is allowed to use the rules of the
analysis of functions of one variable $r$. Especially, one can
multiply these functions and their derivatives of the
corresponding order, one can raise them at various powers, define
functions like $\log$, $\exp$ and other mathematical functions of
$g_{tt}(r)$, $g_{rr}(r)$, $\rho(r)$ and coresponding derivatives.
In general, these operations are forbidden for distributions
\cite{Gelfand}.

In addition, making use of the standard rules for operating with
Dirac $\delta$-function \cite{Gelfand} and accepting (for
simplicity) the following assumptions: 1) the function $\rho(r)$
is a monotonic function of the {\em real} variable $r$; 2)
$\rho(0)=\rho_0$ and the equation $\rho(r)=\rho_0$ has only one
{\em real} solution: $r=0$; one obtains the following alternative
form of the Eq.(\ref{EL}): \ben
\left(\!\sqrt{\!-g_{\rho\rho}}\!-\!{1
\over{\sqrt{\!-g_{\rho\rho}}}}\!\right)\!{d\over{d
\ln\rho}}\!\ln\!\left(\! \rho\left(\!{1\over
{\!-g_{\rho\rho}}}\!-\!1\!\right)\!\right)\!=\nonumber\\ 2
\sigma_0 G_N M_0\,\delta(\rho\!-\!\rho_0),\nonumber\\ {{d^2\ln
g_{tt}}\over{(d \ln\rho})^2}\!+\!{1\over 2}\left(\!{{d\ln
g_{tt}}\over{d \ln\rho}}\!\right)^2\!\!+\!\left(\!1\!+\!{1\over
2}{{d\ln g_{tt}}\over{d \ln\rho}}\!\right){{d\ln
g_{\rho\rho}}\over{d \ln\rho}}\!=\!0, \nonumber\\ {{d\ln
g_{tt}}\over{d \ln\rho}}+g_{\rho\rho}+1\stackrel{w}{=}0.
\la{ELrho}\een

Note that the only remnant of the function $\rho(r)$ in the system
Eq.(\ref{ELrho}) are the numbers $\rho_0$ and
$\sigma_0=sign(\rho^\prime(0))$, which enter only the first
equation, related to the source of gravity. Here $g_{tt}$ and
$g_{\rho\rho}=g_{rr}/ {(\rho^\prime)}^2$ are considered as
functions of the independent variable $\rho \in [\rho_0,\infty)$.

The form of the Eq. (\ref{ELrho}) shows why one is tempted to
consider the Hilbert gauge as a preferable one: In this form the
arbitrary function $\rho(r)$ "disappears".

One usually ignores the general case of an arbitrary value
$\rho_0\!\neq\!0$ accepting, without any justification, the value
$\rho_0=0$, which seems to be natural in Hilbert gauge. Indeed, if
we consider the luminosity distance as a measure of the {\em real}
geometrical distance to the point source of gravity in the 3D
space, we have to accept the value $\rho_0=0$ for the position of
the point source. Otherwise the $\delta$-function term in the Eq.
(\ref{ELrho}) will describe a {\em shell} with radius $\rho_0\neq 0$,
instead of a {\em point} source.

Actually, according to our previous consideration the point source
has to be described using the function  $\delta(r)$. The physical
source of gravity is placed at the point $r=0$ by definition.
There is no reason to change this initial position of the source,
or the interpretation of the variables in the problem at hand. To
what value of the luminosity distance $\rho_0=\rho(0)$ corresponds
the {\em real} position of the point source is not known {\em a
priori}. This depends strongly on the choice of the rho-gauge. One
can not exclude such a nonstandard behavior of the {\em physically
reasonable} rho-gauge function  $\rho(r)$ which leads to some
value $\rho_0 \neq 0$. Physically this means that instead to
infinity, the luminosity of the {\em point} source will go to a
{\em finite} value, when the distance to the source goes to zero.
This very interesting new possibility appears in curved
space-times due to their unusual geometrical properties. It may
have a great impact for physics and deserves further careful
investigation.

In contrast, if one accepts the value $\rho_0=0$, one has to note
that the Hilbert-gauge singularity at $\rho=0$ will be {\em space-like,}
not time-like. This is a quite unusual {\em nonphysical} property
for a physical source of any static physical field.

The choice of the values of the parameters  $\rho_0$ and
$\sigma_0$ was discussed in \cite{Abrams}. There, a new argument
in favor of the choice  $\rho_0=\rho_G$ and $\sigma_0=1$ was given
for different physical problems with spherical symmetry  using an
analogy with the Newton theory of gravity. In contrast, in the present
article we shall analyze the physical meaning of the different
values of the parameter $\rho_0\geq 0$. They turn out to be related
with the gravitational mass defect.

The solution of the subsystem formed by the last two equations of
the system (\ref{ELrho}) is given by the well known functions \ben
g_{tt}(\rho)\!=\!1\!-\!{{\rho_G}/{\rho}},\,\,\,
g_{\rho\rho}(\rho)=-1/g_{tt}(\rho).\la{HilbertSol}\een Note that
in this subsystem one of the equations is a field equation, but
the other one is a constraint. However, these functions do not
solve the first of the Eq.(\ref{ELrho}) for any value of $\rho_0$,
if $M_0\neq 0$. Indeed, for these functions the left hand side of
the first field equation equals identically zero and does not have a
$\delta$-function-type of singularity, in contrast to the
right hand side. Hence, the first field equation remains unsolved.
Thus we see that:

1) Outside the singular point, i.e. for $\rho>\rho_0$, the
Birkhoff theorem is strictly valid and we have a standard form of
the solution, when expressed through the luminosity distance
$\rho$.

2) The assumption that $g_{tt}(r)$, $g_{rr}(r)$ and $\rho(r)$ are
usual $C^3$ smooth functions, instead of distributions, yields a
contradiction, if $M_0\neq 0$. This way one is not able to
describe correctly the gravitational field of a massive point
source of gravity in GR. For this purpose the
first derivative with respect to the variable $r$ at least of one
of the metric coefficients $g_{\mu\nu}$ must have a finite jump,
needed to reproduce the Dirac $\delta$-function in the
energy-momentum tensor of the {\em massive} point particle.

3) The widespread form of the Schwarzschild's solution in Hilbert
gauge (\ref{HilbertSol}) does not describe a gravitational field
of a massive point source and corresponds to the case $M_0=0$. It
does not belong to any physical gauge sector of gravitational
field, created by a massive point source in GR. Obviously, this
solution has a geometrical-topological nature and may be used in
the attempts to reach pure geometrical description of "matter
without matter". In the standard approach to this solution no bare
mass distribution was ever introduced.

The above consideration confirms the conclusion, which was
reached in \cite{ADM} using isotropic gauge for Hilbert solution.
In contrast to \cite{ADM}, in the next Sections we will show how
one can include in GR {\em neutral}
particles with nonzero bare mass $M_0$. The corresponding new solutions
are related to the Hilbert one via singular gauge transformations.

\section{Normal Coordinates for Gravitational Field of a
         Massive Point Particle in General Relativity}

An obstacle for the description of the gravitational field of a point
source at the initial stage of development of GR was the absence
of an adequate mathematical formalism. Even after the development
of the correct theory of mathematical distributions \cite{Gelfand}
there still exist an opinion that this theory is inapplicable to
GR because of the nonlinear character of Einstein equations
(\ref{Einst}). For example, the author of the article \cite{YB}
emphasizes that "the Einstein equations, being non-linear, are
defined essentially, only within framework of functions. The
functionals, introduced in ... physics and mathematics (Dirac's
$\delta$-function, "weak" solutions of partial differential
equations, distributions of Scwartz) are suitable only for linear
problems, since their product is not, in general, defined."  In
the more recent article \cite{GT} the authors have considered
singular lines and surfaces, using mathematical distributions.
They have stressed, that "there is apparently no viable treatment
of point particles as concentrated sources in GR". See also
\cite{Exact} and the references therein. Here we propose a novel
approach to this problem, based on a specific choice of the field
variables.

Let us represent the metric $ds^2$ (\ref{ds0}) of the problem at
hand in a specific form: \ben
\!e^{2\varphi_1}dt^2\!\!-\!e^{\!-2\varphi_1\!+4\varphi_2-2\bar\varphi}dr^2\!\!-\!
\bar\rho^2e^{\!-2\varphi_1\!+2\varphi_2}(d\theta^2\!\!+
\!\sin^2\!\theta d\phi^2) \la{nc} \een where $\varphi_1(r)$,
$\varphi_2(r)$ and $\bar\varphi(r)$ are unknown functions of the
variable $r$ and $\bar\rho$ is a constant -- the unit for
luminosity distance $\rho=\bar\rho\, e^{-\varphi_1+\varphi_2}$. By
ignoring the surface terms in the corresponding integrals, one
obtains the gravitational and the mechanical actions in the form: \ben
{\cal A}_{GR}\!=\!\!{1\over{2 G_N}}\!\!\int\!\!dt\!\!\int\!dr
\Bigl(e^{\bar\varphi}\left(\!-(\bar\rho\varphi_1^\prime)^2\!+
\!(\bar\rho\varphi_2^\prime)^2\right)
\!+\!e^{\!-\bar\varphi}e^{2\varphi_2}\Bigr),\nonumber\\ {\cal
A}_{M_0}\!=-\int\!\!dt\!\!\int\!dr\, M_0\, e^{\varphi_1}\delta(r).
\hskip 3.45truecm \la{A}\een Thus we see that the field variables
$\varphi_1(r)$, $\varphi_2(r)$ and $\bar\varphi(r)$ play the role
of {\em a normal fields' coordinates} in our problem. The field
equations read: \ben  \bar\Delta_r \varphi_1(r)= {\frac
{G_NM_0}{\bar\rho^2}} e^{\varphi_1(r)-\bar\varphi(r)}
\delta(r),\nonumber\\ \bar\Delta_r \varphi_2(r)= {\frac
{1}{\bar\rho^2}}
e^{2\left(\varphi_2(r)-\bar\varphi(r)\right)}\hskip 1.1truecm
\la{FEq}\een where
$\bar\Delta_r\!=\!e^{\!-\bar\varphi}{d\over{dr}} \left(
e^{\bar\varphi} {d \over{dr}} \right)$ is related to the radial
part of the 3D-Laplacean,
$\Delta\!=\!-1/\sqrt{|{}^{(3)}\!g|}\partial_i
\left(\sqrt{|{}^{(3)}\!g|}g^{ij}\partial_j\right)\!=
\!-g^{ij}\left(\partial^2_{ij}-\bar\Gamma_i\partial_j\right)\!=\!\Delta_r
+1/\rho^2\Delta_{\theta\phi} $
:\,\,\,$\bar\Delta_r\!=\!-g_{rr}\Delta_r$.

The variation of the total action with respect to the auxiliary
variable $\bar\varphi$ gives the constraint: \ben
e^{\bar\varphi}\left(-(\bar\rho\varphi_1^\prime)^2\!+\!(\bar\rho\varphi_2^\prime)^2
\right)-e^{-\bar\varphi}e^{2\varphi_2}\stackrel{w}{=}0.
\la{constraint}\een

One can have some doubts about the correctness of the above
derivation of the field equations (\ref{FEq}), because here we
use the Weyl's trick, applying the spherical symmetry directly
to the action functional, not to the Einstein equations
(\ref{Einst}). The correctness of the result of this procedure is
proved in the Appendix A. Therefore, if one prefers, one can
consider the Lagrangian densities (\ref{LGR}), or the actions
(\ref{A}) as an auxiliary tools for formulating of our
one-dimensional problem, defined by the reduced spherically
symmetric Einstein equations (\ref{Einst}), as a variational
problem. The variational approach makes transparent the role, in
the sense of the theory of constrained dynamical systems, of the
various differential equations, which govern the problem.

\section{Regular Gauges and General Regular Solutions of the Problem}

The advantage of the above normal fields' coordinates is that when
expressed through
them the field equations (\ref{FEq}) are linear with respect to
the derivatives of the unknown functions $\varphi_{1,2}(r)$. This
circumstance legitimates the correct application of the mathematical
theory of distributions and makes our normal coordinates privileged
field variables.

The choice of the function $\bar\varphi(r)$ fixes the rho-gauge in the
normal coordinates. We have to choose this function in a way that
makes  the first of the equations (\ref{FEq}) meaningful. Note
that this non-homogeneous equation is quasi-linear and has a
correct mathematical meaning if, and only if, the condition
$|\varphi_1(0)-\bar\varphi(0)|<\infty$ is satisfied.

Let's consider once more the domain $r>0$. In this domain the
first of the equations (\ref{FEq}) gives $\varphi_1(r)=C_1\int
e^{-\bar\varphi(r)}dr+C_2$ with arbitrary constants $C_{1,2}$.
Suppose that the function $\bar\varphi(r)$ has an asymptotics
$\exp(-\bar\varphi(r))\sim k r^n$  in the limit $r\to +0$ (with
some arbitrary constants $k$ and $n$). Then one easily
obtains $\varphi_1(r)-\bar\varphi(r)\sim C_1 k r^{n+1}/(n+1)+
n\ln{r}+\ln{k}+C_2$ -- if $n\neq -1$, and
$\varphi_1(r)-\bar\varphi(r)\sim (C_1k-1)\ln{r} +\ln{k}+C_2$ --
for $n=-1$. Now we see that one can satisfy the condition
$\lim_{r\to 0}|\varphi_1(r)-\bar\varphi(r)|= constant <\infty$ for
arbitrary values of the constants $C_{1,2}$ if, and only if $n=0$.
This means that we must have $\bar\varphi(r)\sim k=const \neq
\pm\infty$ for $r\to 0$. We call such gauges {\em regular
gauges} for the problem at hand. Then $\varphi_1(0)=const\neq
\pm\infty$.  Obviously, the simplest choice of a regular gauge is
$\bar\varphi(r)\equiv 0$. Further on we shall use this {\em basic
regular gauge}. Other regular gauges for the same gauge sector
defer from it by a regular rho-gauge transformation
which describes a diffeomorphism of the
fixed by the basic regular gauge manifold ${\cal
M}^{(3)}\{g_{mn}({\bf r })\}$. In terms of the metric components
the basic regular gauge condition reads $\rho^4
g_{tt}+\bar\rho^4g_{rr}=0$ and gives $\bar\Gamma_r=0$.

Under this gauge the field equations (\ref{FEq}) acquire the
simplest quasi-linear form: \ben \varphi_1^{\prime\prime}(r)=
{\frac {G_NM_0}{\bar\rho^2}}\, e^{\varphi_1(0)}\,
\delta(r),\,\,\,\, \varphi_2^{\prime\prime}(r)= {\frac
{1}{\bar\rho^2}}\,e^{2\varphi_2(r)}.\hskip .2truecm \la{FEq0}\een

The constraint (\ref{constraint}) acquires the simple form: \ben
-(\bar\rho\varphi_1^\prime)^2\!+\!(\bar\rho\varphi_2^\prime)^2
-e^{2\varphi_2}\stackrel{w}{=}0. \la{constraint0}\een

As can be easily seen, the basic regular gauge $\bar\varphi(r)\equiv 0$ has
the unique property to split the system of field equations (\ref{FEq}) and
the constraint (\ref{constraint}) into three independent relations.

The new field equations (\ref{FEq0}) have a general solution \ben
\varphi_1(r)\!=\!{{G_N
M_0}\over{\bar\rho^2}}\,e^{\varphi_1(0)}\bigl(\!\Theta(r)\!-\!\Theta(0)\!\bigr)\,r
\!+\!\varphi_1^\prime(0)\,r\!+\!\varphi_1(0),\nonumber\\
\varphi_2(r)\!=\!-\ln\left({1\over{\sqrt{2\varepsilon_2}}}
\sinh\left(\sqrt{2\varepsilon_2}
{{r_\infty-r}\over{\bar\rho}}\right)\right). \hskip 1.2truecm
\la{Gsol}\een

The first expression in Eq.(\ref{Gsol}) represents a distribution
$\varphi_1(r)$. In it $\Theta(r)$ is the Heaviside step function.
Here we use the {\em additional assumption} $\Theta(0):=1$. It
gives a specific regularization of the products, degrees and
functions of the distribution $\Theta(r)$ and makes them definite.
For example: $\big(\Theta(r)\big)^2=\Theta(r)$,
$\big(\Theta(r)\big)^3=\Theta(r)$, \dots,
$\Theta(r)\delta(r)=\delta(r)$,
$f(r\Theta(r))=f(r)\Theta(r)-f(0)\big(\Theta(r)-\Theta(0)\big)$ --
for any function $f(r)$ with a convergent Taylor series expansion
around the point $r=0$, and so on. This is the only simple
regularization of distributions we need in the present article.

The second expression $\varphi_2(r)$ in Eq.(\ref{Gsol}) is a usual
function of the variable $r$. The symbol $r_\infty$ is used as an
abbreviation for the constant expression
$r_\infty=sign\left(\varphi_2^\prime(0)\right)\bar\rho\,
\sinh\left(\sqrt{2\varepsilon_2}e^{-\varphi_2(0)}\right)/\sqrt{2\varepsilon_2}$.

The constants  \ben \varepsilon_1\!=\!-{1\over
2}\bar\rho^2{\varphi_1^\prime(r)}^2\!+\!{{G_N
M_0}\over{\bar\rho^2}}\varphi_1^{\prime}(0)\,e^{\varphi_1(0)}
\bigl(\!\Theta(r)\!-\!\Theta(0)\!\bigr),\nonumber\\
\varepsilon_2\!=\!{1\over
2}\left(\bar\rho^2{\varphi_2^\prime(r)}^2\!-\!e^{2\varphi_2(r)}\right)
\hskip 3.76truecm \la{epsilon_12}\een are the values of the
corresponding first integrals (\ref{epsilon_12}) of the
differential equations (\ref{FEq0}) for a given solution
(\ref{Gsol}).

Then for the regular solutions (\ref{Gsol}) the
condition(\ref{constraint0}) reads: \ben
\varepsilon_1+\varepsilon_2+{{G_N
M_0}\over{\bar\rho^2}}e^{\varphi_1(0)}
\bigl(\!\Theta(r)\!-\!\Theta(0)\!\bigr)
\stackrel{w}{=}0.\la{constraint_sol} \een

An unexpected property of this relation is that it cannot be
satisfied for any value of the variable $r\in (-\infty,\infty)$,
because $\varepsilon_{1,2}$ are constants. The constraint
(\ref{constraint_sol}) can be satisfied either on the interval
$r\in [0,\infty)$, or on the interval $r\in (-\infty,0)$. If, from
physical reasons we chose it to be valid at only one point
$r^*\in [0,\infty)$, this relation will be satisfied on the whole
interval $r\in [0,\infty)$ and this interval will be the physically
admissible real domain of the radial variable. Thus one can see that
our approach gives a unique possibility {\em to derive} the
admissible real domain of the variable $r$ from the dynamical
constraint (\ref{constraint0}), i.e., this dynamical constraint
yields a geometrical constraint on the values of the radial
variable.

As a result, in the physical domain the values of the first
integrals (\ref{epsilon_12}) are related by the standard equation
\ben\varepsilon_{tot}=\varepsilon_1+\varepsilon_2\stackrel{w}{=}0,
\la{epsilon}\een
which reflects the fact that our variation problem is invariant
under re-parametrization of the independent variable $r$. At the
end, as a direct consequence of the relation (\ref{epsilon}) one
obtains the inequality $\varepsilon_2=-\varepsilon_1>0$, because
in the real physical domain $r\in [0,\infty)$ we have
$\varepsilon_1=-{1\over
2}\bar\rho^2{\varphi_1^\prime(r)}^2=const<0$.

For the function $\rho_{reg,0}(r)\geq 0$, which corresponds to the
basic regular gauge, one easily obtains
\ben\rho_{reg|_{\bar\varphi=0}}(r)=\rho_G
\left(1-\exp\Big({4{{r-r_{\infty}}\over{\rho_G}}}\Big)\right)^{-1}.
\la{reg_rho}\een

Now one has to impose several additional conditions on the
solutions (\ref{Gsol}):

i) The requirement to have an asymptotically flat space-time. The
limit $r\!\to\!r_\infty$ corresponds to the limit
$\rho\!\to\!\infty$. For solutions (\ref{Gsol}) we have the
property $g_{rr}(r_\infty)/{\rho^\prime}^2(r_\infty)\!=\!1$. The
only nontrivial asymptotic condition is $g_{tt}(r_\infty)=1$. It
gives $\varphi_1^\prime(0)\,r_\infty\!+\!\varphi_1(0)\!=\!0$.

ii) The requirement to have the correct Keplerian mass $M$, as seen by
a distant observer.  Excluding the variable $r>0$ from
$g_{tt}(r)=e^{2\varphi_1(r)}$ and $\rho(r)=\bar\rho
e^{-\varphi_1(r)+\varphi_2(r)}$ for solutions (\ref{Gsol}) one
obtains $g_{tt}=1+{{const}\over{\rho}}$, where
$const=2\,sign(\rho)\,sign(\varphi_2^\prime(0))\,\bar\rho^2\,\varphi_1^\prime(0)=-2G_N
M$.

iii) The consistency of the previous conditions with the relation
$g_{tt}\,g_{rr}+{\rho^\prime}^2=0$ gives
$sign(r_\infty)=sign(\rho)=sign(\varphi_1^\prime(0))=sign(\varphi_2^\prime(0))=1$.

iv) The most suitable choice of the unit $\bar\rho$ is
$\bar\rho=G_N M =\rho_G/2$. As a result all initial constants
become functions of the two parameters in the problem --
$r_\infty$ and $M$: $\varphi_1(0)= -{{r_\infty}\over{G_N M}}$,
$\varphi_2(0)= -\ln\left(\sinh\left({{r_\infty}\over{G_N
M}}\right)\right)$, $\varphi_1^\prime(0)= -{{1}\over{G_N M}}$,
$\varphi_2^\prime(0)={{1}\over{G_N
M}}\coth\left({{r_\infty}\over{G_N M}}\right)$.

v) The gravitational defect of the mass of a point particle.

Representing the bare mechanical mass $M_0$ of the point source in
the form $M_0=\int_0^{r_\infty}
M_0\,\delta(r)dr=4\pi\int_0^{r_\infty}
\sqrt{-g_{rr}(r)}\,\rho^2(r)\,\mu(r)dr$, one obtains for the mass
distribution of the point particle the expression
$\mu(r)=M_0\,\delta(r)/\left(4\pi\sqrt{-g_{rr}(r)}\,\rho^2(r)\right)=M_0\,\delta_g(r)$,
where
$\delta_g(r):=\delta(r)/\left(4\pi\sqrt{-g_{rr}(r)}\rho^2(r)\right)$
is the 1D {\em invariant} Dirac delta function. The Keplerian
gravitational mass $M$ can be calculated using the Tolman formula
\cite{books}: \ben M = 4\pi\int_0^{r_\infty}
\rho^\prime(r)\rho^2(r)\mu(r)dr
=M_0\sqrt{g_{tt}(0)}.\la{MKepler}\een Here we use the relation
$\rho^\prime = \sqrt{-g_{tt}\,g_{rr}}$. As a result we reach the
relations:
$g_{tt}(0)\!=\!e^{2\varphi_1(0)}\!=\!\exp\!\left(\!-2{{r_\infty}\over{G_N
M }}\!\right)\!=\!\left(\!{{M}\over{M_0}}\!\right)^2\!\leq\!1$ and
$r_\infty\!=\!G_N M \ln\left({{M_0}\over{M}}\right)\!\geq 0$.
(Note that due to our convention $\Theta(0):=1$ the component
$g_{tt}(r)$ is a continuous function in the interval $r\in
[0,\infty)$ and $g_{tt}(0)=g_{tt}(+0)$ is a well defined
quantity.)

The ratio $\varrho={{M}\over{M_0}}=\sqrt{g_{tt}(0)}\in [0,1]$ describes the
gravitational mass defect of the point particle as a second
physical parameter in the problem. The Keplerian mass $M$ and the
ratio $\varrho$ define completely the solutions (\ref{Gsol}).

For the initial constants of the problem one obtains: \ben
\varphi_1(0)=\ln\varrho ,\,\,\,\varphi_2(0)=- \ln { {1-\varrho^2}
\over {2\varrho} } ,\nonumber\\ \,\nonumber\\ \varphi_1^\prime(0)=
{{1}\over{G_N M}}, \,\,\,\varphi_2^\prime(0)={{1}\over{G_N
M}}\,{{1+\varrho^2}\over{1-\varrho^2}}. \la{In_constMM}\een

Thus we arrive at the following form of the solutions (\ref{Gsol}):

\ben \varphi_1(r)={{r\,\Theta(r)}\over{G_N M}}-\ln(1\!/\!\varrho),
\nonumber\\ \varphi_2(r)\!=\!-\ln\left( {1\over
2}\left({1\!/\!\varrho}\,e^{-r/G_N\!M}-
\varrho\,e^{r/G_N\!M}\right)\right) \la{sol_f} \een

and the rho-gauge fixing function
\ben\rho_{reg|_{\bar\varphi=0}}(r)=\rho_G
\left(1-\varrho^2\exp\Big({4{{r}/{\rho_G}}}\Big)\right)^{-1}.
\la{basic_rho_f}\een

An unexpected feature of this {\em two parametric} variety of
solutions for the gravitational field of a point particle is that each
solution must be considered only in the domain $r\in [\,0,\,G_N
M\ln\left({{1}/{\varrho}}\right)]$, if we wish to have a monotonic
increase of the luminosity distance in the interval
$[\rho_0,\infty)$.

It is easy to check that away from the source, i.e., for $r>0$, these
solutions coincide with the solution (\ref{HilbertSol}) in the
Hilbert gauge. Hence, outside the source the solutions
(\ref{sol_f}) acquire the well known standard form, when
represented using the variable $\rho$. This means that the
solutions (\ref{sol_f}) strictly respect a generalized Birkhoff
theorem. Its proper generalization requires only a justification
of the physical domain of the variable $\rho$. It is remarkable
that for the solutions (\ref{sol_f}) the minimal value of the
luminosity distance is \ben\rho_0=2G_N M /(1-\varrho^2)\geq
\rho_G.\la{rho_0}\een This changes the Gauss theorem and leads to
important physical consequences. One of them is that one must
apply the Birkhoff theorem only in the interval
$\rho\in[\rho_0,\infty)$.

\section{Regular Gauge Mapping of the Interval $r\in [0,r_\infty]$
         onto the Whole Interval $r\in[0,\infty]$}

As we have stressed in the previous section, the solutions
(\ref{sol_f}) for the gravitational field of a point particle must
be considered only in the physical domain $r\in [0,r_\infty]$. It
does not seem to be very convenient to work with such unusual
radial variable $r$. One can easily overcome this problem using
the regular rho-gauge transformation \ben r \to r_\infty{{r/\tilde
r}\over{r/\tilde r+1 }}\la{rrgt}\een with an arbitrary scale
$\tilde r$ of the new radial variable $r$ (Note that in the
present article we are using the same notation  $r$ for different
radial variables.) This linear fractional diffeomorfism does not
change the number and the character of the singular points of the
solutions in the whole compactified complex plane $\tilde{\cal
C}^{(1)}_r$ of the variable $r$. The transformation (\ref{rrgt})
simply places the point $r=r_\infty$ at the infinity $r=\infty$,
at the same time preserving the initial place of the origin $r=0$.
Now the new variable $r$ varies in the standard interval $r\in
[0,\infty)$, the regular solutions (\ref{sol_f}) acquire the final form
\ben \varphi_1(r)=-\ln({1\!/\!\varrho}) \left(1-{{r/\tilde
r}\over{r/\tilde r+1 }}\Theta(r/\tilde r)\right),\nonumber\\
\varphi_2(r)=-\ln\left( {1\over 2
}\left((1\!/\!\varrho)^{{1}\over{r/\tilde
r+1}}-\varrho^{{1}\over{r/\tilde r+1}}\right)\right),\nonumber\\
\bar\varphi(r)=2\ln(r/\tilde r+1)+\ln(\tilde r/r_{\infty}).
\la{sol_NewGauge} \een
The final form of the rho-gauge fixing function reads:
\ben\rho_{reg}(r)=\rho_G
\left(1-\varrho^{{{2}\over{r/\tilde r +1 }}}\right)^{-1}.
\la{rho_NewGauge}\een

The last expression shows that the mathematically admissible
interval of the values of the ratio $\varrho$ is the {\em open}
interval $(0,1)$. This is so, because for $\varrho=0$ and for
$\varrho=1$ we would have impermissible trivial gauge-fixing
functions $\rho_{reg}(r)\!\equiv\!1$ and
$\rho_{reg}(r)\!\equiv\!0$, respectively.

Now we are ready to describe the singular character of the coordinate
transition from the Hilbert form of Schwarzschild solution 
(\ref{HilbertSol})
to the regular one (\ref{sol_NewGauge}).
To simplify notations, let us introduce dimensionless variables
$z=\rho/\rho_g$ and $\zeta=r/\tilde r$.
Then the Eq. (\ref{rho_NewGauge}) shows that the essential part of the
change of the coordinates is described, in both directions,
by the functions:
\ben
z(\zeta)=\left(1-\varrho^{2\over{\zeta+1}}\right)^{-1},\,\,\,
\hbox{and} \nonumber \\
\zeta(z)={{\ln(1/\varrho^2)}\over {\ln z - \ln(z-1)}}-1,
\,\,\,\varrho\in (0,1).
\la{change}
\een

Obviously, the function $z(\zeta)$ is regular at the place of the
point source $\zeta=0$; it has a simple pole at $\zeta=\infty$ and
an essential singular point at $\zeta=-1$. At the same time the
inverse function $\zeta(z)$ has a logarithmic branch points at the
Hilbert-gauge center of symmetry  $z=0$ and at the
``event horizon'' $z=1$.
Thus we see how one produces the Hilbert-gauge
singularities at $\rho=0$ and at $\rho=\rho_{{}_H}$, starting from
a regular solution. The derivative
$$d\zeta/dz={{\ln(1/\varrho^2)}\over{z(z-1)\left(\ln z-\ln(z-1)\right)^2}}$$
aproaches infinity at these two points, hence the singular character of the
change in the whole complex domain of the variables.
Thus we reach a complete description of the change of coordinates and
its singularities in the complex domain of the radial variable.

The restriction of the change of the radial variables on
the corresponding {\em physical} interval outside the source:
$\zeta \in (0, \infty) \leftrightarrow
z\in (1/(1-\varrho^2), \infty)$,
is a regular one.

The expressions (\ref{sol_NewGauge}) and (\ref{rho_NewGauge}) 
still depend on the choice of the units for the new variable $r$. 
We have to fix the arbitrary scale of this variable in the form 
$\tilde r=\rho_G/\ln(1/\varrho^2)=G_NM/\ln\left({{M_0}\over{M}}\right)$ to    
ensure the validity of the standard asymptotic expansion: 
$g_{tt}\sim 1 - \rho_G/r +{\cal O}\left((\rho_G/r)^2\right)$ 
when $r\to\infty$.   

Then the final form of the 4D interval, defined by the new regular 
solutions outside the source is:
\ben
ds^2\!=\!e^{2\varphi_{\!{}_G}}
\left(dt^2\!-\!{{dr^2}\over{N_{\!{}_G}(r)^4}}\right)
\!-\!\rho(r)^2
\left(d\theta^2\!+\!\sin^2\!\theta d\phi^2\right)\!.\hskip .2truecm
\la{New_metric}
\een
Here we are using a modified (Newton-like) gravitational potential:
\ben
\varphi_{\!{}_G}(r;M,M_0):=-{{G_N M}\over {r+G_N M}/\ln({M_0\over M})}
\la{Gpot}
\een
a coefficient 
$N_{\!{}_G}(r)=\left(2\varphi_{\!{}_G}\right)^{-1}
\left(e^{2\varphi_{\!{}_G}}-1\right)$, and 
an optical luminosity distance
\ben
\rho(r)\!=\!2G_NM/\left(1\!-\!e^{2\varphi_{\!{}_G}}\right)
\!=\!{{r\!+\!G_NM/\ln({M_0\over M})}\over {N_{\!{}_G}(r)}}.
\la{rhoG}
\een

These basic formulas describe in a more usual way our regular solutions 
of Einstein equations for  $r\in(0,\infty)$. 
They show immediately that in the limit $\varrho\to 0$ our solutions 
tend to the Pugachev-Gun'ko-Menzel one, and 
$\varphi_{\!{}_G}\left(0;M,M_0\right)=\ln\varrho \to -\infty$. 
In the limit: $\varrho\to 1$ we
obtain for any value of the ratio $r/\rho_g$: 
$g_{tt}(r/\rho_G,\varrho)\to 1$, $g_{rr}(r/\rho_G,\varrho)\to -1$, and
$\rho(r/\rho_G,\varrho)\to \infty$. Because of the last result the 4D
geometry does not have a  meaningful limit when $\varrho\to 1$. 
In this case  
$\varphi_{\!{}_G}\left(r;M,M_0\right) \to 0$ 
at all 3D space points.
Physically this means that solutions without mass defect 
are not admissible in GR.     

\section{Total energy of a point source and its gravitational
field}

In the problem at hand we have an extreme example of an "island
universe``. In it a privileged reference system and a well defined
global time exist. It is well known that under these conditions
the energy of the gravitational field can be defined unambiguously
\cite{books}. Moreover, we can calculate the total energy of the
aggregate of a mechanical particle and its gravitational field.

Indeed, the canonical procedure produces a total Hamilton density
${\cal
H}_{tot}=\Sigma_{a=1,2;\mu=t,r}\,\pi_a^\mu\,\varphi_{a,\mu}-{\cal
L}_{tot}\!=\!{1\over{2G_N}}\left(-\bar\rho^2{\varphi_1^\prime}^2
+\bar\rho^2{\varphi_2^\prime}^2-e^{2\varphi_2}\right)+M_0
e^{\varphi_1}\delta(r)$. Using the constraint (\ref{epsilon}) and
the first of the relations (\ref{In_constMM}), one immediately obtains
for the total energy of the GR universe with one point particle in
it: \ben E_{tot}=\int_0^{{\infty}}{\cal H}_{tot} dr=M=\varrho
M_0\leq M_0\,.\la{E}\een

This result completely agrees with the strong equivalence
principle of GR. The energy of the gravitational field,
created by a point particle is the negative quantity:
$E_{GR}=E_{tot}-E_0=M-M_0=-M_0(1-\varrho)<0$.

The above consideration gives a clear physical explanation of the
gravitational mass defect for a point particle.

\section{Invariant Characteristics of the Solutions with a Point Source}
\subsection{Local Singularities of Point Sources}
Using the invariant 1D Dirac function one can write down the first
of the Eq. (\ref{FEq}) in a form of an {\em exact relativistic}
Poisson equation: \ben \Delta_r \varphi_1(r)=4\pi G_N
M_0\,\delta_g(r).\la{Poisson}\een This equation is a specific
realization of Fock's idea (see Fock in \cite{books}) using our
normal fields coordinates (Sec. V). It can be re-written, too,
in a transparent 3D form:
\ben \Delta \varphi_1(r)=4\pi G_N
M_0\,\delta_g^{(3)}({\bf r}).\la{3DPoisson}\een

The use of the invariant  Dirac $\delta_g$-function has the advantage that
under diffeomorphisms of the space  ${\cal M}^{(3)}\{-g_{mn}({\bf r
})\}$ the singularities of the right hand side of the relativistic
Poisson equation (\ref{Poisson}) remain unchanged. Hence, we have
the same singularity at the place of the source for the whole
class of physically equivalent rho-gauges. Then one can
distinguish the physically different solutions of Einstein
equations (\ref{Einst}) with spherical symmetry by investigating
the asymptotics in the limit $r\to +0$ of the coefficient
$\gamma(r)=1/\left(4\pi\rho(r)^2\sqrt{-g_{rr}(r)}\right)$ in front
of the usual 1D Dirac $\delta(r)$-function in the representation
$\delta_g(r)=\gamma(r)\delta(r)$ of the invariant one.

For regular solutions (\ref{sol_f}), (\ref{sol_NewGauge}) the
limit $r\to +0$ of this coefficient is a constant \ben
\gamma(0)={1\over {4\pi \rho_G^2}} \left({1\over\varrho} -
\varrho\right),\la{gamma}\een which describes the intensity of the
invariant $\delta$-function and leads to the formula
$\delta_g(r)=\gamma(0)\delta(r)$. We see that:

1) The condition $\varrho\in (0,1)$ ensures the correct sign of the
intensity $\gamma(0)$, i.e., the property
$\gamma(0)\in(0,\infty)$, and thus a negativity of the total
energy $E_{GR}$ of the gravitational field of a point particle,
according the previous Section III.

2) For $\varrho\!=\!0$ we have $\gamma(0)\!=\!\infty$, and for
$\varrho\!=\!1$:  $\gamma(0)\!=\!0$. Hence,  we have non-physical
values of the intensity $\gamma(0)$ in these two cases. The
conclusion is that {\em the physical interval} of values of the
ratio $\varrho$ is the open interval  $(0,1)$. This is consistent
with the mathematical analysis of this problem, given in the
previous Section.

It is easy to obtain the asymptotics of the coefficient $\gamma(r)$
in the limit $r\to 0$ for other solutions, considered in the
Introduction:
\\

Schwarzschild solution: \hskip 0.65truecm $\gamma_{{}_S}(r)\sim
{1\over {4\pi \rho_G^2}}\left({{\rho_G}\over r}\right)^{1/2}$;

Hilbert solution: \hskip 1.58truecm $\gamma_{{}_H}(r)\sim {i\over
{4\pi \rho_G^2}}\left({{\rho_G}\over r}\right)^{5/2}$;

Droste solution: \hskip 1.68truecm $\gamma_{{}_D}(r)\sim {1\over
{4\pi \rho_G^2}}$;

Weyl solution: \hskip 1.95truecm $\gamma_{{}_W}(r)\sim {16\over
{\pi \rho_G^2}}\left({r\over{\rho_G}}\right)^4$\,\,;

Einstein-Rosen solution: \hskip 0.4truecm $\gamma_{{}_{ER}}(r)\sim
{1\over {4\pi \rho_G^2}}\left({{\rho_G}\over r}\right)^{1/2}$;

Isotropic (t-r) solution: \hskip 0.6truecm
$\gamma_{{}_{ER}}(r)\sim {1\over {4\pi \rho_G^2}}\left({{x}\over
{1-x}}\right)^{1/2}$;

Pugachev-Gun'ko-Menzel

\hskip 2.1truecm solution: \hskip 0.5truecm
$\gamma_{{}_{PGM}}(r)\!\sim\!{1\over {4\pi
\rho_G^2}}\!\left(\!{{r}\over
{\rho_G}}\!\right)^{\!2}\!\!e^{\rho_G/2r}$.
\\

As we can see:

1) Most of the listed solutions are physically different. Only two
of them: Schwarzschild and Einstein-Rosen ones, have the same
singularity at the place of the point source and, as a result,
have diffeomorphic spaces ${\cal M}^{(3)}\{g_{mn}({\bf r })\}$,
which can be related by a regular rho-gauge transformation.

2) As a result of the alteration of the physical meaning of the
variable $\rho=r$ inside the sphere of radius $\rho_{{}_H}$, in
Hilbert gauge the coefficient $\gamma(r)$ tends to {\em
imaginary} infinity for $r\to 0$. This is in a sharp contrast to
the {\em real} asymptotic of all other solutions in the limit
$r\to 0$.

3) The Droste solution is a regular one and corresponds to a
gravitational mass defect with
$\varrho_{{}_D}=\left(\sqrt{5}-1\right)/2\approx 0.61803$. The
golden-ratio-conjugate number $\left(\sqrt{5}-1\right)/2$, called
sometimes also "a silver ratio", appears in our problem as a root
of the equation $1/\varrho-\varrho=1$, which belongs to the
interval $\varrho\in (0,1)$. (The other root of the same equation
is $\,-\!\left(\sqrt{5}+1\right)/2<0$ and does not correspond to a
physical solution.) Hence, the Droste solution can be transformed
to the form (\ref{sol_f}), or (\ref{sol_NewGauge}) by a proper
regular rho-gauge transformation.

4) The Weyl solution resembles a regular solution with
$\varrho_{{}_W}=1$ and $\gamma_{{}_W}(0)=0$. Actually the exact
invariant $\delta$-function for this solution may be written in
the form (see the Appendix B): \ben
\delta_{g,{}_W}(r)={{16}\over{\pi \rho_G^2}} \left(
\Big({{r}\over{\rho_G}}\Big)^4\delta(r)+
\left({{\rho_G}\over{r}}\right)^4\delta\Big({{\rho_G^2}\over
r}\Big) \right).\la{Wdelta}\een This distribution acts as a
zero-functional on the standard class of smooth test functions
with compact support, which are finite at the points $r=0,\infty$.

The formula (\ref{Wdelta}) shows that:

a) The Weyl solution describes a problem with a spherical symmetry
in the presence of {\em two} point sources: one with
$\varrho_{{}_W}^0=1$ at the point $r=0$ and another one with
$\varrho_{{}_W}^\infty=1$ at the point $r=\infty$. Hence, it turns
out that this solution is the first {\em exact} analytical
two-particle-like solution of Einstein equations (\ref{Einst}).
Unfortunately, as a two-point solution, the Weyl solution is
physically trivial: it does not describe the dynamics of two point
particles at a finite distance, but rather gives only a static
state of two particles at an infinite distance.

b) To make these statements correct, one has to compactify the
comformally flat 3D space ${\cal M}^{(3)}\{-g_{mn}({\bf r })\}$
joining it to the infinite point $r=\infty$. Then both the Weyl
solution {\em and} its source will be invariant with respect to
the inversion $r \rightarrow \rho_G^2/r$.

5) The isotropic (t-r) solution resembles a regular one, but in it
$x\approx 1.27846$ is the only real root of the equation
$x+\ln(x-1)=0$ and gives a non-physical value of the parameter
$\varrho_{{}_I}\approx 2.14269$, which does not belong to the
physical interval $(0,1)$.

6) The singularity of the coefficient $\gamma(r)$ in front
of the usual 1D Dirac $\delta$-function in the Pugachev-Gun'ko-Menzel
solution is stronger than any polynomial ones in the other listed solutions.

\subsection{Local Geometrical Singularities of the Regular
          Static Spherically Symmetric Space-Times}
It is well known that the scalar invariants of the Riemann tensor
allow a manifestly coordinate independent description of the
geometry of space-time manifold. The problem of a single point
particle can be considered as an extreme case of a perfect fluid,
which consists of only one particle. According to the article
\cite{CMcL}, the maximal number of independent {\em real}
invariants for such a fluid is 9. These are the scalar curvature $R$
and the standard invariants $r_1, w_1, w_2, m_3, m_5$. (See
\cite{CMcL} for their definitions.) The invariants $R$, $r_1$ and
$m_3$ are real numbers and the invariants $w_1$,$w_2$ and $m_5$
are, in general, complex numbers.  The form of these invariants
for spherically symmetric space-times in normal field
variables is presented in Appendix C. Using these formulas we
obtain for the regular solutions (\ref{sol_f}) in the basic gauge
the following simple invariants:

\ben
I_1\!=\!{1\over{8\rho_G^2}}{{(1\!-\!\varrho^2)^4}\over{\varrho^2}}
\delta\left(\!{r\over{\rho_G}}\!\right)\!=\!{{\pi(1\!-\!\varrho^2)^3}\over{2\varrho}}
\delta_g\left(\!{r\over{\rho_G}}\!\right)\!=\!-{1\over 2}R(r),
\hskip .truecm \nonumber
\\ I_2=0,\hskip 7.6truecm \nonumber\\
I_3\!=\!{{\Theta(r/\rho_G)\!-\!1}\over{8\rho_G^2\varrho^2}}\left(1\!-\!\varrho^2
e^{4r/\rho_G}\right)^4\!=
\!{{\rho_G^2}\over{8\varrho^2}}{{\Theta(r/\rho_G)\!-\!1}\over{\rho(r)^4}},\hskip
.7truecm\nonumber\\
I_4\!=\!{{\Theta(r/\rho_G)}\over{4\rho_G^2}}\left(1\!-\!\varrho^2
e^{4r/\rho_G}\right)^3\!=
\!{{\rho_G}\over{4}}{{\Theta(r/\rho_G)}\over{\rho(r)^3}}. \hskip
1.8truecm
\la{inv_sol}\een

As can be seen easily:

1) The invariants $I_{1,...,4}$ of the Riemann tensor are well
defined distributions. This is in accordance with the general
expectations, described in the articles \cite{TaubRaju}, where one
can find a correct mathematical treatment of distributional valued
curvature tensors in GR.

As we saw, the manifold ${\cal M}^{(1,3)}\{g_{\mu\nu}(x)\}$ for
our regular solutions has a geometrical singularity at the point,
where the physical massive point particle is placed.

Note that the metric for the regular solutions
is globally continuous, but its first
derivative with respect to the radial variable $r$ has a finite jump
at the point source. This jump is needed for a correct
mathematical description of the delta-function distribution of
matter of the massive point source in the right hand side of
Einstein equations (\ref{Einst}).

During the calculation of the expressions $I_{3,4}$ we have used
once more our assumption $\Theta(0)=1$.

Three of the invariants (\ref{inv_sol}) are independent on the
real axes $r\in (-\infty,\infty)$ and this is the true number of
the independent invariants in the problem of a single point source
of gravity. On the real physical interval $r\in [0,\infty)$ one
has $I_3=0$ and we remain with only two independent invariants.
For $r\in (0,\infty)$ the only independent invariant is $I_4$, as
is well known from the case of Hilbert solution.

2) The scalar invariants $I_{1,...,4}$ have the same form as in
Eq. (\ref{inv_sol}) for the regular solutions in the
representation (\ref{sol_NewGauge}), obtained via the
diffeomorphic mapping (\ref{rrgt}).

3) All other geometrical invariants $r_1(r)$, $w_{1,2}(r)$,
$m_{3,5}(r)$ include degrees of Dirac $\delta$-function and are
not well defined distributions. Therefore the choice of the simple
invariants $I_{1,...,4}$ is essential and allows us to use
solutions of Einstein equations, which are distributions.

To the best of authors knowledge, the requirement to make possible
the correct use of the theory of distributions is a novel criteria
for the choice of invariants of the curvature tensor and seems to
not have been used until now. It is curious to know if it is
possible to apply this new criteria in more general cases as
opposed to the case of static spherically symmetric space-times.

4) The geometry of the space-time depends essentially on both
parameters $M$ and $\varrho$, which define the
regular solutions (\ref{sol_NewGauge}) of Einstein
equations (\ref{Einst}). Hence, for different values of the two
parameters these solutions describe non-diffeomorphic
space-times with different geometry.

\subsection{Event Horizon for Static Spherically Symmetric Space-Times
with Point Singularity}

According to the well known theorems by Hawking, Penrose, Israel
and many other investigators, the only solution with a regular
event horizon not only in GR, but in the theories with scalar
field(s) and in more general theories of gravity, is the Hilbert
one \cite{FN}.  As we have seen, this solution has a pure
geometrical nature and does not describe a gravitational field of
a point particle with bare mechanical mass $M_0\neq 0$. The strong
mathematical results like the well known no hear theorems, etc,
for the case of metrics with an event horizon can be shown to be
based on to the assumption that such horizon is {\em indeed
present} in the solution. These mathematical results do not
contradict to the ones, obtained in the present article, because
the other solutions, which we have considered together with the
Hilbert one, do not have an event horizon at all.

Indeed, for the point  $\rho_{{}_H}$ at which
$g_{tt}(\rho_{{}_H})=0$ one obtains $\rho(\rho_{{}_H})=\rho_G$.
The last equations do not have any solution $\rho_{{}_H}\in {\cal
C}^{(1)}$ for the regular solutions (\ref{sol_f}),
(\ref{sol_NewGauge}). The absence of a horizon in the physical 
domain $r\in [0,\infty)$ is obvious in the representation (\ref{New_metric})
of the regular solutions.

For the other classical solutions one obtains in the limit $r\to
\rho_{{}_H}$ as follows:\\

Schwarzschild solution:\hskip .65truecm  $g_{rr}(\rho_{{}_H})\sim
-{{\rho_G}\over{r}}$, \,\,\,\, $\rho_{{}_H}=0$;

Hilbert solution: \hskip 1.35truecm $g_{rr}(\rho_{{}_H})\!\sim\!
-{{\rho_G}\over{r-\rho_G}}$, \, $\rho_{{}_H}\!=\!\rho_G$;

Droste solution: \hskip 1.5truecm $g_{rr}(\rho_{{}_H})\sim -1$,
\,\, \,\,\,\,\,\,$\rho_{{}_H}=0$;

Weyl solution: \hskip 1.7truecm $g_{rr}(\rho_{{}_H})\sim -1$, \,\,
\,\,\,\,\,$\rho_{{}_H}=\rho_G$;

Einstein-Rosen solution: \hskip 0.2truecm $g_{rr}(\rho_{{}_H})\sim
-{{\rho_G}\over{r}}$,\,\,\,\,\,\, $\rho_{{}_H}=0$ ;

Isotropic (t-r) solution: \hskip 0.4truecm
$g_{rr}(\rho_{{}_H})\sim e^{{r}\over{\rho_G}}$, \,\,
$\rho_{{}_H}\!=\!-\infty$;

Pugachev-Gun'ko-Menzel

\hskip 2.1truecm solution: \hskip 0.truecm $g_{rr}(r)\!\sim\!
\!-\!\left({{\rho_G}\over{r}}\right)^4e^{-{{r}\over{\rho_G}}}$,\,
$\rho_{{}_H}=0$.
\\

As one can see, only for Hilbert, Schwarzschild and Einstein-Rosen
solutions the metric component $g_{rr}(r)$ has a simple pole at
the point $\rho_{{}_H}$. In the last two cases this is
a point in the corresponding manifold ${\cal M}^{(3)}\{g_{mn}({\bf r })\}$,
where the center of symmetry is placed, not a horizon.


\section{On the Geometry of Massive Point Source in GR}

It is curious that for a metric, given by a regular solution, the
3D-volume of a ball with a small radius $r_b<\rho_G$, centered at
the source, is $$Vol(r_b)={4\over 3} \pi \rho_G^3
{{12\varrho}\over{(1-\varrho^2)^4}} {{r_b}\over{\rho_G}}+ {\cal
O}_2({{r_b}\over{\rho_G}}).$$ It goes to zero linearly with
respect to $r_b\to 0$, in contrast to the Euclidean case, where
$Vol_E(r_b)\sim r_b^3$. This happens, because for the regular
solutions (\ref{sol_f}), (\ref{sol_NewGauge}) we obtain
$\sqrt{|{}^3g|}=\rho(r)^2\sin\theta \to \rho_0\sin\theta \neq 0$
in the limit $r \to 0$. Nevertheless, $lim_{r_b\to 0}\,
Vol(r_b)=0$ and this legitimates the use of the term "a point
source of gravity" in the problem at hand:
the source can be surrounded by a sphere with an
arbitrary small volume $Vol(r_b)$ in it and with an arbitrary
small radius $r_b$.

In contrast, when $r_b\to 0$ the area of the ball's surface has a
finite limit: ${{4\pi \rho_G^2}\over {(1-\varrho^2)^2}}> 4\pi
\rho_G^2$, and the radii of the big circles on this surface tend
to a finite number ${{2\pi \rho_G}\over{1-\varrho^2}}>2\pi
\rho_G$.

Such unusual geometry, created by the massive point sources in GR,
may have an interesting physical consequences. For example, space-times,
defined by the regular solutions (\ref{sol_f}),
(\ref{sol_NewGauge}), have an unique property: When one approaches
the point source, its luminosity remains finite. This leads to a
very important  modification of Gauss theorem in the corresponding
3D spaces. After all, this modification may solve the well known
problem of the {\em classical} divergences in field theory,
because, as we see, GR offers a natural cut-off parameter 
$\tilde r=\rho_G/\ln(1/\varrho^2)$ for
fields, created by massive point particles. If this will be
confirmed by more detailed calculations, the price, one must pay
for the possibility to overcome the old
classical divergences problem will be to accept the idea, that the
point objects can have a nonzero surface, having at the same time
a zero volume and a zero size $r$.

On the other hand, as we have seen,  the inclusion of point
particles in GR is impossible without such unusual geometry. The
Hilbert solution does not offer any more attractive alternative
for description of a source of gravity as a physical massive {\em
point}, placed  at the coordinate point $\rho=0$, and  with {\em
usual} properties of a geometrical point. Hence, in GR we have no
possibility to introduce a notion of a physical point particle
with nonzero bare mass without a proper change of the standard
definition for point particle as a mathematical object, which has
simultaneously zero size, zero surface and zero volume. We are
forced to remove at least one of these three usual requirements as
a consequence of the concentration of the bare mass $M_0$ at only one
3D space point.

From physical point of view the new definition for physical point
particle, at which we arrived in this article, is obviously
preferable.

Moreover, it seems natural to have an essential
difference between the geometry of  the ``empty'' space-time
points and that of the matter points in GR. The same geometry for
such essentially different objects is possible only in the
classical physics, where the space-time geometry does not depend
on the matter. Only under the last assumption there is no {\em
geometrical} difference between pure mathematical space points and
matter points. The intriguing new situation, described in this
article, deserves further careful analysis.

The appearance of the above nonstandard geometry in the point mass
problem in GR was pointed at first by Marcel Brillouin \cite{Brillouin}.
Our consideration of the regular solutions (\ref{sol_f}),
(\ref{sol_NewGauge}) confirms his point of view on the character
of the singularity of the gravitational field of massive point
particles in GR.


\section{Concluding Remarks}

The most important result of the present article is the explicit
indication of the fact that there exist infinitely many different static
solutions of Einstein equations (\ref{Einst}) with spherical symmetry,
a point singularity, placed at the center of symmetry,
and vacuum outside this singularity, and with the same Keplerian mass $M$.
These solutions fall into different gauge classes, which
describe physically and geometrically different space-times.
Some of them were discovered at the early stage
of development of GR, but up to now they are often considered as
equivalent representations of some "unique" solution which depends on 
only one parameter -- the Keplerian mass $M$.
As shown in Section IX A, this is not the case.

As a consequence of Birkhoff theorem, when expressed in the same variables,
the solutions with the same mass $M$ indeed coincide in their common 
regular domain of coexistence -- outside the singularities. 
In contrast, they may have a different behavior at the corresponding 
singular points. 

A correct description of a massive point source of gravity is impossible,
making use of most of these classical solutions.

Using novel normal coordinates for the gravitational field of a single
point particle with bare mechanical mass $M_0$ we are able to
describe correctly the massive point source of gravity in GR. The
singular gauge transformations yield the possibility to overcome
the restriction to have a zero bare mass $M_0$ for neutral point
particles in GR \cite{ADM}. It turns out that this problem has a
two-parametric family of regular solutions.

One of the parameters -- the bare mass $M_0$ of the point source,
can be obtained in a form of surface integral, integrating both
sides of Eq. (\ref{3DPoisson}) on the whole 3D space ${\cal
M}^{(3)}\{-g_{mn}({\bf r})\}$: \ben M_0={1\over{4 \pi
G_{{}_N}}}\int_{{\cal M}^3}d^3{\bf r} \sqrt{|{}^3g|}\,\Delta
\left(\ln\sqrt{g_{tt}}\right)=\nonumber \\ {1\over{4 \pi
G_{{}_N}}}\oint_{\partial{\cal M}^3}d^2\sigma_i
\sqrt{|g|}\,\,g^{ij}\,\partial_j {\sqrt{g^{tt}}}.\nonumber\een

The second parameter -- the  Keplerian mass $M$, is described,
as shown in Section VIII, by the total energy:
$$M=\int_{{\cal M}^3}d^3{\bf r}\sqrt{|{}^3g|}\,{\cal H}_{tot}=
\int_{{\cal M}^3}d^3{\bf r}\sqrt{|g|}\,M_0\, \delta_g^{(3)}({\bf r}).$$

These two parameters define the {\em gauge class} of the given
regular solution. Obviously the parameters $M$ and $M_0$ are invariant
under regular static gauge transformations, i.e., under diffeomorphisms
of the 3D space ${\cal M}^{(3)}\{-g_{mn}({\bf r})\}$. It is an analytical
manifold with a strong singularity at the place of the massive point 
particle.
For every regular solution both parameters are finite and positive and
satisfy the additional physical requirement $0<M<M_0$.

It is convenient to use a more physical set of continuous
parameters for fixing the regular solutions, namely:
the Keplerian mass $M\in (0,\infty)$ and the  gravitational
mass defect ratio $\varrho={{M}\over{M_0}}\in(0,1)$.

The only classical solution, which is regular, is the Droste
one. For this solution the gravitational mass defect ratio is
$\varrho_{{}_D}=\left(\sqrt{5}-1\right)/2$.

For the regular solutions the physical values of the optical
luminosity distance $\rho$ are in the semi-constraint interval
$\rho\in [{{\rho_G}\over{1-\varrho^2}}, \infty)$.

Outside the source, i.e., for $\rho>{{\rho_G}\over{1-\varrho^2}}$,
the Birkhoff theorem is strictly respected for all regular
solutions.

The metric, defined by a regular solution, is {\em globally}
continuous, but its first derivatives have a jump at the point
source. Our explicit results justify the Raju's conclusion
\cite{TaubRaju} that GR will be consistent with the existence of point
particles if one assumes the metric to be at most of class $C^0$
and show that the Hawking-Penrose singularity theory,
based on the assumption that the components of $g_{\mu\nu}$ are
{\em at least} $C^1$ functions, must be reconsidered.

For the class of regular solutions, written in the form (\ref{sol_f}),
(\ref{sol_NewGauge}), or (\ref{New_metric}), 
the non-physical interval of the optical
luminosity distance $\rho\in[0,{{\rho_G}\over{1-\varrho^2}}]$,
which includes the luminosity radius $\rho_{{}_H}=\rho_G$, can be
considered as an "optical illusion". All pure mathematical
objects, which belong to this interval, are in the imaginary
domain "behind" the real physical source of the gravitational
field and have to be considered as a specific kind of optical
"mirage". This is in agreement with Dirac's conclusion about the
non-physical character of the inner domain $\rho\leq \rho_G$ for
the Hilbert solution \cite{Dirac} and extends this conclusion on
the whole non-physical interval of the optical luminosity distance
$\rho\in[0,{{\rho_G}\over{1-\varrho^2}}]>\rho_{{}_H}$ for a given
{\em regular} solution.

We are forced to cut the Hilbert form of the Schwarzschild
solution at the value $\rho_0={{\rho_G}\over{1-\varrho^2}}$
because of the presence of the matter point mass. Its presence
ultimately requires a definite jump of the first derivatives of
the metric. The jump is needed to make the Einstein tensor
coherent with the energy-momentum stress tensor of the point
particle. This cutting can be considered as a further development
of Dirac's idea \cite{Dirac}:

{\em ``Each particle must have a finite
size no smaller than the Schwarzschild radius. I tried for some
time to work with a particle with radius equal to the
Schwarzschild radius, but I found great difficulties, because the
field at the Schwarzschild radius is so strongly singular, and it
seems that a more profitable line of investigation is to take a
particle bigger than the Schwarzschild radius and to try to
construct a theory for such particle interacting with
gravitational field.''}

For a precise understanding of these statements one has to take
into account that the above Dirac's idea is expressed in terms
of the luminosity distance $\rho$.

An alternative development, which is much closer to Dirac approach 
to the above problem, can be found in the recent articles \cite{MM}.  

The geometry around physical massive point particles is
essentially different from the geometry around the "empty"
geometrical points. This unusual geometry, described at first by
Brillouin, may offer a new way for overcoming of the {\em
classical} fields divergence problems. The new possibility differs
from the one, used in \cite{ADM} for charged point particles. It
is based on the existence of a natural cut-off parameter 
$\tilde r=G_{\!{}_N}M\left(c^2\ln\left({{M_0}\over M}\right)\right)^{-1}$ 
and on a new interpretation of the  relation between the
singular mass distribution of a point particle and the geometry of
the space-time around this particle.

According to the remarkable comment by Poincar\'e \cite{Poincare}
real problems can never be considered as solved or unsolved,
but rather they are always {\em more or less} solved.

The strong {\em physical} singularity at the ``event horizon'' of
the Hilbert solution, stressed by Dirac and further physical
consequences, which one can derive from the new
regular solutions (\ref{sol_NewGauge}) of the old
problem, considered here will be discussed in a separate article.
In this one we will add some more remarks.

Our consideration shows that the observed by the astronomers
compact dark objects (CDO), called by them black {\em holes}
without any direct evidences for the existence of {\em real} 
event horizons \cite{AKL}, 
can not be described theoretically by solutions
(\ref{HilbertSol}) in Hilbert gauge, if we assume that these
objects are made from matter of some nonzero bare mechanical mass
$M_0\neq 0$. At present the only real fact is the existence of a massive
invisible CDO with Keplerian mass $M$, which is too large with respect to
the conventional understanding of the stars physics.

Concerning these unusual CDO, most probably one actually has to
solve a much more general problem. Since a
significant amount of invisible dark matter, which manifests
itself only due to its gravitational field, is observed in the
Nature at very different scales: in the clusters of galaxies, in
the halos of the galaxies, at the center of our galaxy, and  as a
compact dark components in some binary star systems, one is
tempted to look for some universal explanation of all these
phenomena. Obviously such universal explanation can not be based
on Schwarzschild solution in Hilbert gauge and one has to look for
some other theoretical approach. A similar idea was pointed out
independently in the recent articles \cite{RN}.

At the end we wish to mention some additional open problems, both
mathematical and physical ones, connected with the gravitational
field of point particles.

Our consideration was essentially restricted to the real domain
of variables. The only exception was the description of the
singular change (\ref{change}) of radial variable.
It is well known that the natural domain
for study of the solutions of holomorphic differential equations
is the complex one. A complete knowledge of the
solutions of such equations is impossible without description of
all singularities of the solutions in the complex domain.
Therefore, looking for a complete analysis of the solutions of
some differential equation of $n$-th order
$f\left(w^{[n]}(z),w^{[n-1]}(z),...,w^{[1]}(z),w(z),z\right)=0$
for a function $w(z), z\in {\cal C}^{(1)}$, where $f(...)$ is a
holomorphic function of the corresponding complex variables, one
has to consider these solutions as a holomorphic functions of
complex variable $z$. Now the most important issue becomes the
study of the singular points of the solutions in the {\em whole}
complex domain.

For the Einstein's equations (\ref{Einst}) such a {\em four-dimensional}
complex analysis is impossible at present, because the corresponding
mathematical methods are not developed enough. But for the problem
of single point source in GR, in its 1D formulation, used in this article,
one can use the classical complex analysis.
Moreover, using the well known complex
representation of distributions (see Bremermann in \cite{Gelfand}),
it is not difficult to generalize the classical results for
the case when the differential equation has on the right hand side
a distribution, which depends on the variable $z$.
Such a term may lead to a discontinuity of the solution $w(z)$,
or of its derivatives.
The corresponding complex analysis of the solution remains
an important open mathematical problem, together with the problem
of finding and classifying {\em all} static solutions of Einstein
equations with one point singularity and spherical symmetry.

The most important physical problem becomes to find criteria for
an experimental and/or observational probe of different solutions
of Einstein equations with spherical symmetry. Since there exist
{\em physical and geometrical} differences between the solutions
of this type, one can find a real way of distinguishing them
experimentally. Especially, a problem of the present day is to
answer the question about which one of the spherically symmetric
solution gives the right description of the observed CDO, if one
hopes that the CDO can be considered in a good approximation as
spherically symmetric objects with some Keplerian mass $M$ and some 
mass defect ratio $\varrho\in (0,1)$. 
In this case, according to the Birkhoff theorem, 
we must use the new solutions (\ref{New_metric}) for description of 
their gravitational field in the outer vacuum domain, outside the CDO. 

On the same reason the solutions (\ref{New_metric}) must be used, too, 
for description of the vacuum gravitational field of 
spherically symmetric bodies of finite size, 
made of usual matter,  outside their radius $R$. 
Because of the specific character
of the correction of the Newton's low of gravitation, 
defined by the term $G_{\!{}_N}Mc^{-2}/\ln\left({{M_0}\over M}\right)$ in 
the potential $\varphi_{\!{}_G}(r;M,M_0)$ (\ref{Gpot}),  
the difference between our solutions  (\ref{New_metric}) and 
the Hilbert one is in higher order relativistic terms, when 
$\ln\left({{M_0}\over{M}}\right)\sim 1$. 
This difference will not influence the rho-gauge invariant local 
effects, like perihelion shift, deviation of light rays, 
time-dilation of electromagnetic pulses, and so on, but it may be 
essential for the quantities, which depend on the precise form of 
the potential $\varphi_{\!{}_G}(r;M,M_0)$ (\ref{Gpot}). 
For all bodies in the solar system the term 
$G_{\!{}_N}Mc^{-2}/\ln\left({{M_0}\over M}\right)$ is very small in 
comparison with their radius $R$: 
$\left(\rho_{\!{}_G}/2R\right)_{Earth}\sim 10^{-10}$,  
$\left(\rho_{\!{}_G}/2R\right)_{\odot }\sim 10^{-6}$, etc. 
Nevertheless, the small differences between the Newton's potential
and the modified one may be observed in the near future 
in the extremely high precision measurements  under programs like 
APOLLO, LATOR, etc. (see for example \cite{LATOR}), if we will 
be able to find a proper quantities, which depend on the precise form of 
the potential $\varphi_{\!{}_G}(r;M,M_0)$ (\ref{Gpot}). 
A better possibility for observation of a deviations from Newton's  
gravitational potential and in four-interval may offer neutron stars, since 
for them $\left(\rho_{\!{}_G}/2R\right)_{n*}\sim 0.17-0.35$.

The results of the present article may have an important impact on
the problem of gravitational collapse in GR, too. Up to now most
of the known to the author investigations presuppose to have as a
final state of the collapsing object the Hilbert
solution, or a proper generalization of it with some kind of event
horizon, in the spirit of the cosmic censorship
hypothesis. In the last decade more attention was paid to
the final states with a necked singularities. The
existence of a two-parametric class of regular static solutions
without event horizon opens a new perspective for these
investigations. Most probably the proper understanding of the origin
of the mass defect of point particle, introduced in the present
article, will be reached by a correct description of the
gravitational collapse.

Without any doubts, the inclusion of the new static spherically
symmetric regular solutions of Einstein equations in the corresponding
investigations will open new perspectives for a further
developments in GR, as well as in its modern generalizations.

\begin{acknowledgments}

I'm deeply indebted: to prof. Ivan Todorov and to the participants
of his seminar for illuminating discussions, comments and
suggestions, to prof. Salvatore Antoci, to Dr.~James Brian Pitts,
and to prof. Lluis Bel for bringing important references to my
attention and for some useful suggestions and comments. My special
thanks  to prof. Salvatore Antoci for sending me copies of
articles  \cite{Dirac}. I wish to thank too to prof. Vasil Tsanov
for useful discussion about multi-dimensional complex analysis,
to prof. G.~Schaefer and to prof. S. Bonazzola for useful
discussions and comments on the basic ideas of present article,
to Dr.~Roumen Borissov for proofreading the manuscript and to
Dr.~Nedjalka Petkova for her help in translation of the article
\cite{Brillouin}.

This research was supported in part  by the Fulbright Educational
Exchange Program, Grant Number 01-21-01, by Scientific Found of
Sofia University, Grant Number 3305/2003, by Special Scientific
Found of JINR, Dubna for 2003, by the University of Trieste, by INFN,
and by the Abdus Salam International  Centre for Theoretical Physics,
Trieste, Italy.
\end{acknowledgments}

\appendix

\section{Direct derivation of Field Equations from Einstein Equations}

Using the normal field variables, introduced in Section V, one
can write down the nonzero mixed components $G^\mu_\nu$ of
Einstein tensor in the basic regular gauge ($\bar\varphi(r)\equiv
0$) as follows: \ben G^t_t\!=\!
{{e^{2\varphi_1-4\varphi_2}}\over{\bar\rho^2}}
\Big(2\bar\rho^2\varphi_1^{\prime\prime}\!-\!2\bar\rho^2\varphi_2^{\prime\prime}
\!-\!(\bar\rho\varphi_1^\prime)^2\!+\!(\bar\rho\varphi_2^\prime)^2\!+\!e^{2\varphi_2}
\Big),\nonumber\\
G^r_r\!=\!{{e^{2\varphi_1-4\varphi_2}}\over{\bar\rho^2}}
\Big(\!(\bar\rho\varphi_1^\prime)^2\!-\!(\bar\rho\varphi_2^\prime)^2\!+
\!e^{2\varphi_2}\Big),
\nonumber\hskip 2.5truecm\\ G^\theta_\theta=
G^\phi_\phi=-{{e^{2\varphi_1-4\varphi_2}}\over{\bar\rho^2}}
\Big(\bar\rho^2\varphi_2^{\prime\prime}\!+\!(\bar\rho\varphi_1^\prime)^2\!-
\!(\bar\rho\varphi_2^\prime)^2\Big).\hskip 1.truecm
\la{Gmunu}\een

Taking into account that the only nonzero component of the
energy-momentum tensor is $T^t_t$ and combining the Einstein
equations (\ref{Einst}), one immediately obtains the field
equations (\ref{FEq0}) and the constraint (\ref{epsilon}). A
slightly more general consideration with an arbitrary rho-gauge
fixing function $\bar\varphi(r)$ yields the equations (\ref{FEq})
and the constraint(\ref{constraint}).

A similar derivation produces the equations (\ref{EL}), which
present another version of the field equations in spherically
symmetrical space-times. In this case: \ben
G^t_t\!=\!{1\over{-g_{rr}}}\left(\!-2\left(\!{{\rho^\prime}\over{\rho}}\!\right)^{\!\prime}
\!-\!3\left(\!{{\rho^\prime}\over{\rho}}\!\right)^{\!2}\!+
\!2{{\rho^\prime}\over{\rho}}{{\sqrt{-g_{rr}}^{\,\prime}}\over{\sqrt{-g_{rr}}}}
\right)\!+\!{{1}\over{\rho^2}},\hskip .3truecm \nonumber\\
G^r_r\!=\!{1\over{-g_{rr}}}\left(\!-\!\left(\!{{\rho^\prime}\over{\rho}}\!\right)^{\!2}\!+
\!2{{\rho^\prime}\over{\rho}}{{\sqrt{g_{tt}}^{\,\prime}}\over{\sqrt{g_{tt}}}}
\right)\!+\!{{1}\over{\rho^2}},\hskip 2.2truecm  \\
G^\theta_\theta\!=\! G^\phi_\phi\!=\!{1\over{-g_{rr}}}
\Bigg(\!\!-\!\left(\!{{\rho^\prime}\over{\rho}}\!\right)^{\!\prime}\!
\!-\!\left(\!{{\rho^\prime}\over{\rho}}\!\right)^{\!2}\!
\!-\!{{\rho^\prime}\over{\rho}}{{\sqrt{g_{tt}}^{\,\prime}}\over{\sqrt{g_{tt}}}}
\!+\hskip 1.4truecm \nonumber
\\
\!{{\rho^\prime}\over{\rho}}{{\sqrt{-g_{rr}}^{\,\prime}}\over{\sqrt{-g_{rr}}}}\!-\!
\left(\!{{\sqrt{g_{tt}}^{\,\prime}}\over{\sqrt{g_{tt}}}}\!\right)^{\!\prime}\!
\!-\!\left(\!{{\sqrt{g_{tt}}^{\,\prime}}\over{\sqrt{g_{tt}}}}\!\right)^{\!2}\!
\!+\!{{\sqrt{g_{tt}}^{\,\prime}}\over{\sqrt{g_{tt}}}}
{{\sqrt{-g_{rr}}^{\,\prime}}\over{\sqrt{-g_{rr}}}}
\Bigg).\nonumber \la{Gmunu_g}\een

\section{Invariant Delta Function for Weyl Solution}

The function $\rho_{{}_W}(r)\!=\!{{1}\over
4}\!\left(\!\sqrt{r/{\rho_G}}\!+\!\sqrt{{\rho_G}/
r}\right)^2\!\geq\!\rho_G$ which describes the Weyl rho-gauge
fixing is an one-valued function of $r\in (0,\infty)$, but its
inverse function $r_{{}_W}(\rho)$ is a two-valued one:
$r_{{}_W}(\rho)=r_{{}_W}^\pm(\rho)=\rho\left(1\pm\sqrt{1-\rho_G/\rho}\right)^2$,
$\rho\in[\rho_G,\infty)$. Obtaining information about distant
objects only in optical way, i.e. by measuring only the luminosity
distance $\rho$, one is not able to choose between the two
branches of this function and therefore one has to consider both
of them as possible results of the observations. Hence, the
3D space ${\cal M}^{(3)}_{{}_W}\{g_{mn}({\bf r })\}$, which
appears in the Weyl gauge, has to be considered as a two-sheeted
Riemann surface with a branch point at the 2D sphere
$\rho^*=\rho_G$ in ${\cal M}^{(3)}_{{}_W}\{g_{\tilde m \tilde
n}({\rho,\tilde \theta,\tilde \phi})\}$ -- the "observable" space
with new spherical coordinates ${\rho,\tilde \theta,\tilde \phi}$.
Because of the limits $r_{{}_W}^+(\rho)\to \infty$ and
$r_{{}_W}^-(\rho)\to 0$ for $\rho\to \infty$, one can write down
the invariant $\delta$-function in the case of Weyl gauge in the
form \ben
\delta_{g,{}_W}(\rho)={{\sqrt{1-\rho_G/\rho}}\over{4\pi}\rho^4}\,
\delta\!\left({1\over\rho}\right).\la{W_delta_rho}\een

Now, taking into account that the equation $1/\rho(r)=0$ has two
solutions: $r=0$ and $1/r=0$ and using the standard formula for
expansion of the distribution  $\delta\big(1/\rho(r)\big)$, we
obtain easily the representation (\ref{Wdelta}).

\section{Independent Nonzero Invariants of Riemann Curvature in Normal Field 
Coordinates}
Using the the representation (\ref{nc}) of the metric in normal
field variables after some algebraic manipulations one obtains the following
expressions for the possibly independent nonzero invariants of
Riemann curvature tensor in the problem at hand:

\ben R=-2e^{2(\varphi_1-2\varphi_2)}
\left(\varphi_1^{\prime\prime}-2E_2-E_3 \!\right),\hskip
2.6truecm\nonumber\\ r_1\!=\!{1\over
4}e^{4(\varphi_1\!-\!2\varphi_2)}
\Bigg(\!2\Big(\varphi_1^{\prime\prime}\!-\!E_2\!-\!E_3\Big)^2\!+\!
\Big(\varphi_1^{\prime\prime}\!+\!E_3\Big)^2\Bigg),\hskip
.4truecm\nonumber\\ w_1\!={{C^2}\over {6}},\,\,\,
w_2\!=-{{C^3}\over {36}},\hskip 4.7truecm \la{invars}\\
m_3\!=\!{{C^2}\over{12}}e^{4(\varphi_1\!-\!2\varphi_2)}
\Bigg(\!\Big(\varphi_1^{\prime\prime}\!-\!E_2\!-\!E_3\Big)^2\!\!+\!
2\Big(\varphi_1^{\prime\prime}\!+\!E_3\Big)^2\Bigg),\hskip
.15truecm\nonumber\\
m_5\!=\!{{C^3}\over{36}}e^{4(\varphi_1\!-\!2\varphi_2)}
\Bigg(\!{1\over 2}\!
\Big(\varphi_1^{\prime\prime}\!-\!E_2\!-\!E_3\Big)^2\!\!-\!
2\Big(\varphi_1^{\prime\prime}\!+\!E_3\Big)^2\Bigg),\nonumber\een
where $E_2:=\varphi_2^{\prime\prime} - e^{2\varphi_2}/\bar\rho^2$
and
$E_3:=\varphi_1^\prime{}^2-\varphi_2^\prime{}^2+e^{2\varphi_2}/\bar\rho^2$.

As we see, the invariants $w_{1,2}$ and $m_5$ in our problem are
real. Hence, in it we may have at most 6 independent invariants.

In addition $(w_1)^3=6\,(w_2)^2$ and the metric (\ref{nc}) falls
into the class II of Petrov classification \cite{CMcL}. As a
result the number of the independent invariants is at most 5.

For our purposes it is more suitable to use as an independent
invariant \ben C:=-e^{2\varphi_1\!-\!4\varphi_2}
\Big(2\varphi_1^{\prime\prime}\!+\!
6\varphi_1^{\prime}\left(\varphi_1^{\prime}\!-\!\varphi_2^{\prime}\right)
\!-\!E_2\!-\!2E_3\Big) \la{C}\een instead of the invariants $w_1$ and
$w_2$. The invariant $C$ has the following advantages:

i) It is linear with respect of the derivative
$\varphi_{1}^{\prime\prime}$. This property makes meaningful the
expression (\ref{C}) in the cases when
$\varphi_{1}^{\prime\prime}$ is a distribution;

ii) It is a homogeneous function of first degree with respect to
the Weyl conformal tensor $C_{\mu\nu,\lambda}{}^\xi$;

iii) It is proportional to an eigenvalue of a proper tensor, as
pointed, for example, in Landau and Lifshitz \cite{books}.

iv) The difference between the scalar curvature $R$ and  the
invariant (\ref{C}) yields a new invariant \ben
D\!:=\!R\!-\!C\!=\!3\, e^{2\varphi_1\!-\!4\varphi_2}
\Big(2\varphi_1^{\prime}\big(\varphi_1^{\prime}\!-\!\varphi_2^{\prime}\big)
+E_2 \Big)\,\,\,\,\la{D}\een which does not include the derivative
$\varphi_1^{\prime\prime}$. Due to this property the invariant $D$
in the case of regular solutions  (\ref{sol_f}),
(\ref{sol_NewGauge}) will not contain a Dirac $\delta$-function.
One can use the new invariant $D$ as an independent one, instead
of the invariant $C$.

As seen from Eq. (\ref{invars}), the following functions of 
the invariants $r_1$, $m_3$ and
$m_5$:
\ben \, e^{2(\varphi_1\!-\!2\varphi_2)}
\Big(\varphi_1^{\prime\prime} +E_3
\Big)=\sqrt{4\left(m_3/w_1-r_1\right)/3},\hskip .35truecm
\nonumber
\\
e^{2(\varphi_1\!-\!2\varphi_2)}\Big(2\varphi_1^{\prime\prime}\!-\!E_2\!-\!E_3
\Big)\!=\!\sqrt{2\left(4r_1\!-\!m_3/w_1\right)/3}\hskip .5truecm
\la{m5}\een are linear with respect to the derivative
$\varphi_1^{\prime\prime}$. Using proper linear combinations of
these invariants and the scalar curvature $R$ one can easily check
that the quantities $E_2\,e^{2(\varphi_1\!-\!2\varphi_2)}$, and
$E_3\,e^{2(\varphi_1\!-\!2\varphi_2)}$ are independent invariants
too.

Thus, our final result is that, in general, for the metric
(\ref{nc}) in the basic regular gauge $\bar\varphi\equiv 0$ the
Riemann curvature tensor has the following four independent
invariants: \ben
I_1:=e^{2(\varphi_1\!-\!2\varphi_2)}\varphi_1^{\prime\prime},\hskip
3.77truecm\nonumber\\
I_2:=e^{2(\varphi_1\!-\!2\varphi_2)}\left(\varphi_2^{\prime\prime}-
e^{2\varphi_2}/\bar\rho^2\right),\hskip 1.73truecm\\
I_3:=e^{2(\varphi_1\!-\!2\varphi_2)} \left(-\varphi_1^\prime{}^2+
\varphi_2^\prime{}^2-e^{2\varphi_2}/\bar\rho^2
\right)/2,\nonumber\\
I_4:=e^{2(\varphi_1\!-\!2\varphi_2)}\varphi_1^{\prime}
\big(\varphi_1^{\prime}\!-\!\varphi_2^{\prime}\big)/2. \hskip
2.03truecm\nonumber \la{I}\een These invariants are linear with
respect to the second derivatives of the functions $\varphi_{1,2}$
-- a property, which is of critical importance when we have to
work with distributions $\varphi_{1,2}^{\prime\prime}$.

\end{document}